\documentclass[a4paper,oneside]{amsart}
\usepackage{graphics}
\usepackage{amssymb}
\usepackage{amsmath}
\usepackage{geometry}
\usepackage{hyperref}
\hypersetup{breaklinks=true}
\newcommand{\Ret}{\mathrm{Ret}}
\newcommand{\Fun}{\mathrm{Fun}}
\newcommand{\LL}{\mathcal{L}}
\newcommand{\PP}{\mathcal{P}}
\newcommand{\DD}{\mathcal{D}}
\newcommand{\MM}{\mathcal{M}}
\newcommand{\OO}{\mathcal{O}}
\newcommand{\TT}{\mathcal{T}}

\newcommand{\Det}{\mathrm{det}}
\newcommand{\tr}{\mathrm{tr}}
\newcommand{\dd}{\partial}
\newcommand{\ad}{\mathrm{ad}}
\newcommand{\Gr}{\mathrm{Gr}}
\newcommand{\gh}{\mathrm{gh}}
\newcommand{\N}{\mathcal{N}}
\newcommand{\F}{\mathcal{F}}
\newcommand{\BV}{\mathrm{BV}}
\newcommand{\g}{\mathfrak{g}}
\newcommand{\G}{\mathcal{G}}
\newcommand{\e}{\epsilon}
\newcommand{\RR}{\mathbb{R}}
\newcommand{\ZZ}{\mathbb{Z}}
\newcommand{\TTT}{\mathbf{T}}
\newcommand{\LLL}{\mathbf{L}}
\newcommand{\Str}{\mathrm{Str}}
\newcommand{\Aut}{\mathrm{Aut}}

\newcommand{\ev}{\mathrm{ev}}
\newcommand{\id}{\mathrm{id}}
\newcommand{\vol}{\mathrm{vol}}
\newcommand{\Iter}{\mathrm{Iter}}
\newcommand{\Loop}{\mathrm{Loop}}
\newcommand{\A}{\mathcal{A}}
\newcommand{\B}{\mathcal{B}}
\newcommand{\bt}{\bullet}
\newcommand{\odd}{\mathrm{odd}}
\newcommand{\even}{\mathrm{even}}
\newcommand{\I}{\mathcal{I}}

\newcommand{\be}{\begin{equation}}
\newcommand{\ee}{\end{equation}}

\newtheorem{Lemma}{Lemma}
\newtheorem{Theorem}{Theorem}

\begin{document}

\author{P. Mn\"ev}
\address{PDMI RAS, 27, Fontanka, 191023, Saint-Petersburg, Russia}
\email{pmnev@pdmi.ras.ru}
\title{Notes on simplicial $BF$ theory}

\thanks{Supported by grants: RFBR 05-01-00922, RAS presidium program ``mathematical problems of nonlinear
dynamics'', NSh-8065.2006.2, CRDF grant RUM1-2622-ST-04}
\begin{abstract}
In this work we discuss the construction of ``simplicial $BF$
theory'', the field theory with finite-dimensional space of fields,
associated to a triangulated manifold, that is in a sense equivalent
to topological $BF$ theory on the manifold (with
infinite-dimensional space of fields). This is done in framework of
simplicial program --- program of constructing discrete topological
field theories. We also discuss the relation of these constructions
to homotopy algebra.
\end{abstract}
\maketitle

\setcounter{tocdepth}{2} \tableofcontents
\section{\bf Introduction}
This work contains first results obtained by the author in framework
of simplicial program proposed to him by Andrei Losev, also some
preliminary arguments are included. The idea of simplicial program
is to take some topological quantum field theory $\TT_M$ on manifold
$M$ and formulate discrete field theory $\TT_\Xi$ on triangulation
$\Xi$ of $M$, which is (in some sense) equivalent to $\TT_M$. Having
succeeded in constructing $\TT_\Xi$, we can use it to compute
quantities associated to $\TT_M$, like state sum or vacuum
expectation values of observables, in terms of finite-dimensional
integrals (since space of fields of $\TT_\Xi$ is
finite-dimensional), instead of functional integrals. Thus, having a
discrete theory $\TT_\Xi$, we leave behind all subtleties of
functional integrals, like regularization and possibility of
renormalization.

To our knowledge the simplicial program was successfully completed
for abelian Chern-Simons theory (see \cite{Adams}, \cite{SSSA}).

We consider a special topological theory, ``extended $BF$ theory'',
which is the ordinary non-abelian $BF$ theory with fields promoted
to extended fields --- non-homogeneous differential forms (also
called ``super fields'' in physical literature). From physical point
of view, extended $BF$ theory is ordinary $BF$ theory with ghosts,
ghosts for ghosts etc. taken into account, and arises naturally in
process of quantization. This theory perfectly fits into
Batalin-Vilkovisky formalism as a special case of what we call
``abstract $BF$ theory''. The latter is a class of theories
associated to differential graded Lie algebras, for extended $BF$
case --- associated to algebra of differential forms on manifold $M$
with values in a gauge Lie algebra $\g$. The action of extended $BF$
theory is a generating function for structure constants of
commutator and differential on the algebra of $\g$-valued
differential forms, and Batalin-Vilkovisky master equation is
equivalent to the three quadratic relations for forms: $d^2=0$,
Leibniz identity and Jacobi identity.

For any abstract $BF$ theory, associated to some DGLA $\G$, if $\G$
is decomposed into sum of two subcomplexes $\G=\G'\oplus\G''$ with
$\G''$ acyclic, we construct induced (in physical terminology,
``effective'') theory, associated to $\G'$. The action of this
induced theory is constructed via an integral ``over $\G''$'' (more
precisely, an integral over Lagrangian submanifold in
$\Pi\G''\oplus[\G'']^*$) --- the Batalin-Vilkovisky integral. The
question of convergence of the latter is quite subtle for
infinite-dimensional $\G''$; we hope that it is at least
perturbatively well-defined, and our calculations of induced action
for $BF$ theory on simplex confirm this conjecture (but do not prove
it in full generality, of course). The tree part of effective action
is the generating function for $L_\infty$ algebra operations on
$\G'$, and quadratic relations on the operations follow from
classical master equation. The general theorem states that $\G'$ and
$\G$ are quasi-isomorphic as $L_\infty$ algebras. In terms of BV
integral, the quasi-isomorphism between $\G'$ and $\G$ is explicitly
constructed via ``expectation value map'' (theorem \ref{L_infty
morphism}).

By Koszul duality, to define $L_\infty$ structure on $\G'$ is the
same as to define cohomological vector field $Q$ (i.e. an odd vector
field satisfying $Q^2=0$) on parity-reversed space $\Pi\G'$. The
1-loop part of effective action defines a $Q$-invariant measure
(volume form) on $\Pi\G'$. The $Q$-invariance follows from quantum
master equation. And there are no higher-loop contributions to BV
integrals of abstract $BF$ theory. Thus there are classical (given
by tree approximation) effects in BV integral
--- the induced $L_\infty$ operations on $\G'$ and the
quasi-isomorphism $\G'\rightarrow\G$, and quantum effect (given by
1-loop contributions) --- the $Q$-invariant measure on $\Pi\G'$.
Calculation of the quantum effect is much more involved: values of
1-loop Feynman diagrams for effective action are expressed as
certain super traces over $\G''$ and may contain divergencies if
$\G''$ is infinite-dimensional. The induced theory associated to
$\G'$ is in some sense ``homotopic'' to initial theory associated to
$\G$. By ``homotopic'' we mean that $L_\infty$ structure on $\G'$
generated by tree part of effective action is quasi-isomorphic to
DGLA structure on $\G$. Whether the notion of homotopy of $L_\infty$
algebras can be extended to $L_\infty$ algebras with $Q$-invariant
measure, is an interesting question.

We use the name ``classical higher operations'' for terms of Taylor
expansion of tree part of effective action --- these correspond to
$L_\infty$ operations on $\G'$, and name ``quantum higher
operations'' for terms of Taylor expansion of 1-loop part of
effective action (which is the logarithm of density of $Q$-invariant
measure on $\Pi\G'$).

For the sake of simplicial program we are interested in constructing
induced theory for extended $BF$ theory on manifold $M$, associated
to subcomplex of $\g$-valued differential forms, consisting of
$\g$-valued Whitney forms on triangulation $\Xi$ of $M$. We call
this induced theory ``simplicial $BF$ theory'' on $\Xi$. From
general arguments, this theory on triangulation, with
finite-dimensional space of fields, is ``homotopic'' to extended
$BF$ theory on $M$.

Technically, to write down simplicial $BF$ action for any
triangulation $\Xi$, it is sufficient to solve the problem for
single simplex in each dimension (this property is formulated as the
factorization of BV integral for simplicial action on triangulation
--- theorem \ref{factorization thm}). Thus only one universal calculation is needed in each
dimension. In dimension $D=0$ it is trivial, for $D=1$ --- not quite
trivial, but can be done exactly, and explicit formula for
simplicial $BF$ action on 1-simplex is written (theorem \ref{thm:
reduced action on 1-simplex}). For dimensions $D>1$ we do not know
closed expression for effective action on $D$-simplex, but we have
computed first classical and quantum higher operations (values of
simplest Feynman diagrams for BV integral for the effective action
are computed). We would like to emphasize that calculation of
effective action on simplex is absolutely universal and done once
and for all time --- having it, we have defined simplicial $BF$
theory on any triangulation of any manifold, and may conduct further
calculations starting from this discrete topological field theory,
via finite-dimensional BV integrals.

One immediate use of simplicial $BF$ theory is as follows. We can
construct induced $BF$ theory on de Rham cohomologies of manifold
$M$. The tree part of effective action is the generating function
for Massey operations on cohomologies, and 1-loop part provides a
$Q$-invariant measure on parity-reversed space of cohomologies. But
now instead of calculating this effective action via functional BV
integral, starting from extended $BF$ theory on $M$, we may induce
theory on cohomologies from simplicial $BF$ theory on triangulation
$\Xi$, via finite-dimensional integral. The 1-loop part of effective
action on cohomologies may be integrated to give the state sum of
extended $BF$ theory, which therefore again can be computed from
simplicial $BF$ theory, avoiding functional integrals.

Other possible uses of simplicial $BF$ theory include the
construction of knot (and higher-dimensional knot) invariants in
terms of vacuum expectation values of certain observables in
simplicial $BF$ theory. Another possible use is combinatorial
construction of characteristic classes. We intend to elaborate on
these points in the future.

\subsection{Main results}
\begin{itemize}
\item The fact that simplicial $BF$ action on triangulation $\Xi$ decomposes into sum over simplices of $\Xi$ of
some local contributions, which we call ``reduced effective
actions'' on simplices (theorem \ref{factorization thm}). This
statement is the reason why calculation of simplicial $BF$ action on
one simplex in each dimension is universal and allows to define
simplicial $BF$ action on any triangulation.

\item Explicit expression for reduced effective action on 1-simplex (theorem \ref{thm: reduced action on
1-simplex}), obtained by direct computation of corresponding BV
integral. This result allows us to fully construct simplicial $BF$
theory on 1-dimensional simplicial complexes. We also use it to
illustrate simplicial approach by computing state sum of $BF$ theory
on circle $Z(S^1)$ starting from simplicial $BF$ action on
discretized circle and calculating finite-dimensional BV integral.
We show that $Z(S^1)$ equals the volume of gauge group. The
expression for reduced effective action on 1-simplex is also an
important ingredient for defining simplicial $BF$ action in
dimensions $D>1$.

\item Explicit expressions for first classical higher operations on
simplex of arbitrary dimension (theorem \ref{thm: first classical
higher operations}).

\item Result of direct calculation of simplest non-trivial quantum
operation $q^{(2)}$ for 2-simplex and for 3-simplex (theorem
\ref{operation 2->0 in D=2, D=3}). This computation is quite long
and contains divergent quantities in intermediate stages, and thus
requires regularization. Not quite surprisingly, the final answers
are finite.

\item Partial result for first quantum operation $q^{(2)}$ on simplex of
arbitrary dimension: symmetry allows only two possible terms for
$q^{(2)}$. We recover coefficient for one of the terms from quantum
master equation and known classical higher operations (theorem
\ref{thm: first quantum operation}), while the other term is
$Q$-exact, and thus the corresponding coefficient cannot be
recovered in this way. This result agrees with theorem
\ref{operation 2->0 in D=2, D=3} in dimensions 2 and 3, but is much
cheaper, in a sense that it does not require hard calculations.
\end{itemize}

Another important general construction (not specific to the
simplicial setting) is the construction of $L_\infty$
quasi-isomorphism between DGLA (or more generally, $L_\infty$
algebra) $\G$ and induced $L_\infty$ structure on its subcomplex
$\G'$ via BV integral (theorem \ref{L_infty morphism}).

\subsection{Open problems}
The following is the beginning of long list of interesting questions
one can ask about simplicial $BF$ theory:
\begin{itemize}
\item Question of Wilson renormalization of simplicial $BF$ action.
If $\Xi'$ is a triangulation and $\Xi$ is some subdivision of
$\Xi'$, we can induce effective action $\tilde{S}_{\Xi'}$ on $\Xi'$
from simplicial $BF$ action $S_\Xi$ on $\Xi$. The question is: does
$\tilde{S}_{\Xi'}$ obtained in this way differ from simplicial $BF$
action $S_{\Xi'}$ on $\Xi'$ (obtained by standard induction from
extended $BF$ theory on manifold), and if yes, what is the
difference? General arguments indicate that this difference
(``renormalization'') should be BV-exact, in a sense that
exponentials $e^{\tilde{S}_{\Xi'}/\hbar}$ and $e^{S_{\Xi'}/\hbar}$
belong to the same cohomology class of BV-Laplacian. We checked that
in dimension $D=1$ there is no such renormalization, while already
for $D=2$ the first quantum operation gets renormalized under
barycentric aggregation of triangulation (while first classical
higher operations are not renormalized).

\item More general setting for the previous question is as follows.
Let $\G$ be a DGLA and $\Ret_\G$ be the category of retracts, where
objects are subcomplexes $\G'\subset \G$, containing all cohomology
of $\G$ and morphisms are retractions. As we explain in this paper,
information contained in induced $BF$ theory ($BF_\infty$ theory) on
subcomplex $\G'\in\Ret_\G$ is equivalent to the pair $(Q,\rho)$
where $Q$ is a cohomological vector field on $\Pi\G'$ and $\rho$ is
the $Q$-invariant measure on $\Pi\G'$. Then operation of induction
of $BF_\infty$ theory from $\G'$ to $\G''\subset\G'$ with Lagrangian
manifold for BV integral defined by given chain homotopy operator
$K$, may be regarded as ``parallel transport'' of
$(Q,\rho)$-structure on objects of $\Ret_\G$ along morphism $K$.
Then general setting for question about Wilson renormalization is:
what can we say about holonomy of this parallel transport?

\item Problem of constructing observables for simplicial $BF$
theory. Particularly we are interested in observables, associated to
knots and higher dimensional knots. These should be some discrete
analogs of observables constructed in \cite{CCRFM}.

\item Simplicial $BF$ action defines curvature of a discretized
superconnection. A natural question is: how to use it to write local
combinatorial formulae for characteristic classes?

\item Question of relation between state-sum of extended $BF$ theory
on a manifold (calculated via simplicial $BF$ theory on
triangulation) and Turaev-Viro-type invariants of manifolds,
calculated as sum over colorings of triangulation (see \cite{TV}).

\item Extension of our simplicial constructions to Poisson sigma
model (see \cite{SchStr}), and their application to deformation
quantization and Kontsevich integrals (see \cite{Kontsevich},
\cite{CF}).
\end{itemize}

\subsection{Sources and literature}
The simplicial program for topological field theories was inspired
by problem of constructing combinatorial version of Chern-Simons
theory. This problem was proposed by M. Atiyah in \cite{Atiyah}. Our
main sources for geometric interpretation of Batalin-Vilkovisky
formalism are \cite{AKSZ} and \cite{Schwarz}. Our source on
infinity-algebras is \cite{MSS}. One of key constructions for our
treatment of simplicial $BF$ theory
--- the construction of Dupont's chain homotopy between differential forms on
manifold $M$ and Whitney forms, associated to triangulation of $M$,
is borrowed from \cite{Getzler}. The construction of effective
action for $BF$ theory is explained in \cite{KLG}. In unpublished
paper \cite{Kozak} an alternative treatment of simplicial program is
given. Paper \cite{Schatz} explains induction of effective action on
cohomologies on tree level in mathematical rigor.

\subsection{Acknowledgements}
I wish to thank my advisor L.D. Faddeev for support and discussion,
and Andrei Losev for inspiration. The setting of problem and many
ideas here are due to discussions with A. Losev. I also wish to
thank Nikolai Mn\"ev, Nikolai Gromov, Andrei Kozak and Florian
Sch\"atz for useful discussions and comments.

\section{\bf Extended $BF$ theory in Batalin-Vilkovisky formalism}
Ordinary $BF$ theory on $D$-dimensional manifold $M$ with compact
gauge group $G$ is defined by classical action \be
S(\omega,B)=\tr\int_M B\wedge F \label{BF action}\ee
 where $F=d\omega+\omega\wedge \omega$ is the curvature
of connection 1-form $\omega$ on $M$ with values in Lie algebra $\g$
of group $G$, and $B$ is $(D-2)$-form on $M$ with values in $\g$.
Trace is taken in some representation of $\g$.

\subsection{Extended $BF$ theory: fields, action}
Now we move to the extended $BF$ theory. Let
$\Omega(M)=\Omega^0(M)\oplus\cdots\oplus\Omega^D(M)$ be the
commutative differential graded algebra (cDGA) of differential forms
on $M$ (with $\Omega^k(M)$ being the subspace of $k$-forms)--- the
de Rham algebra with de Rham differential and wedge product. We
denote by $\G=\g\otimes\Omega(M)=\G^0\oplus\cdots\oplus\G^D$ the
differential graded Lie algebra (DGLA) of $\g$-valued differential
forms on $M$. Let $\{e_\alpha\}$ be some basis in $\G$ and
$\{e^\alpha\}$ be the dual basis in $\G^*$. We also suppose that
each basis element $e_\alpha$ is homogeneous and denote its degree
by $|\alpha|$, which means that $e_\alpha\in\G^{|\alpha|}$. Parity
of $e_\alpha$ (and parity of $e^\alpha$) is equal to parity of
integer $|\alpha|$.

The extended $BF$ action is defined as\be
S(\omega,p)=<p,d\omega+\frac{1}{2}[\omega,\omega]>\label{extended BF
action}\ee where $<\bullet,\bullet>$ denotes the canonical pairing
of $\G$ and $\G^*$. The fields $\omega$ and $p$ are
\begin{eqnarray}
\omega=\sum_{\alpha}\omega^\alpha e_\alpha \\
p=\sum_{\alpha}e^\alpha p_\alpha
\end{eqnarray}
where $\{\omega^\alpha\}$ are  variables of parity opposite to
parity of $|\alpha|$, while $\{p_\alpha\}$ are variables of parity
coinciding with parity of $|\alpha|$. Thus $\omega$ is totally odd
and $p$ is totally even. Field $\omega$ belongs to space
$\overline{\Pi\G}$, the totally odd version of $\G$ (which is no
longer a DGLA, but just a vector super space):
$$\omega\in\overline{\Pi\G}:=[\RR^{1|1}\otimes\G]^\odd=
\Pi\G^0\oplus\G^1\oplus\Pi\G^2\oplus\cdots\oplus\Pi^{D+1}\G^D$$
while $p$ belongs to $\overline{\G^*}$, the totally even version of
$\G^*$:
$$p\in\overline{\G^*}:=[\G^*\otimes\RR^{1|1}]^\even
=[\G^0]^*\oplus\Pi[\G^1]^*\oplus[\G^2]^*\oplus\cdots\oplus\Pi^D[\G^D]^*$$
Here $\Pi$ is the parity reversing operation on vector super spaces,
$\Pi^k=\Pi$ for $k$ odd and $\Pi^k=\id$ for $k$ even. We also use
the traditional notation $\RR^{k|l}=\RR^k\oplus\Pi\RR^l$, and for a
vector super space $X$ we denote its even and odd subspaces by
$[X]^\even$ and $[X]^\odd$ respectively.

The action
\begin{multline}
S(\omega,p)=<p,d\omega+\frac{1}{2}[\omega,\omega]>=\\=
\sum_{\alpha,\beta}
(-1)^{|\beta|+1}<e^\alpha,d\,e_\beta>p_\alpha\,\omega^\beta+
\frac{1}{2}\sum_{\alpha,\beta,\gamma}
(-1)^{|\beta|\,(|\gamma|+1)}<e^\alpha,[e_\beta,e_\gamma]>
p_\alpha\,\omega^\beta \omega^\gamma \label{BF action in
coordinates}
\end{multline}
belongs to the space $\RR[[\{\omega^\alpha\},\{p_\alpha\}]]$ of
formal power series of variables $\{\omega^\alpha\}$ and
$\{p_\alpha\}$. This space is an associative commutative super
algebra freely generated by variables $\{\omega^\alpha\}$ and
$\{p_\alpha\}$. Thus we may think of it as of algebra of functions
$\Fun(\F)$ on space $\F=\Pi\G\oplus\G^*$. The latter is called
``space of fields''. To match more general constructions of
Batalin-Vilkovisky formalism (see \cite{AKSZ}), we may identify $\F$
with cotangent bundle to $\Pi\G$ with reversed parity in fibers:
\be\F=\Pi T^*(\Pi\G)\label{space of fields}\ee In this picture
$\omega^\alpha$ are coordinate functions on base of $\F$, while
$p_\alpha$ are coordinate functions on fibers of $\F$. Our initial
fields $\omega$ and $p$ are then ``generating functions'' for
coordinate functions on base and on fibers of $\F$ respectively,
useful to write short formulae for various objects of extended $BF$
theory (as in (\ref{BF action in coordinates})).

Less formally, extended $BF$ action (\ref{extended BF action}) is
obtained from (\ref{BF action}) by promoting $\omega$ and $B$ to
extended fields (also called ``super fields'') and introducing a new
field $p$, related to $B$ by lowering an index:
$<p,\bullet>=\tr\int_M B\wedge\bullet$. Field $\omega$ is a
non-homogeneous differential form on $M$ with values in gauge
algebra $\g$:
$$\omega=\omega^{(0)}+\cdots+\omega^{(D)}$$ with $\omega^{(k)}$
being $\g$-valued $k$-form with parity opposite to parity of integer
$k$. Field $p$ is decomposed as $$p=p^{(0)}+\cdots+p^{(D)}$$ with
$p^{(k)}$ a $k$-coform (element of $[\Omega^{k}(M)]^*$, or
equivalently $(D-k)$-form with lowered index) with values in
coalgebra $\g^*$ and with parity equal to parity of $k$. Thus
$\omega$ is a non-homogeneous $\g$-valued form of total parity $1$,
while $p$ is a non-homogeneous $\g^*$-valued coform of total parity
$0$.

\subsubsection*{\bf Remark} One can give an alternative description of
extended $BF$ theory in terms of $\ZZ\oplus\ZZ$ grading: we may
prescribe ``ghost numbers'' to $\{\omega^\alpha\}$ and
$\{p_\alpha\}$ instead of just parities:
$\gh(\omega^\alpha)=1-|\alpha|$, $\gh(p_\alpha)=-2+|\alpha|$. Then
we say that fields $\omega$ and $p$ belong to spaces
$[\Gr\otimes\G]^{\deg+\gh=+1}$ and $[\G^*\otimes\Gr]^{\deg+\gh=-2}$
respectively, where $\Gr=\bigoplus_{k=-\infty}^\infty \RR^{[+k]}$
--- the vector super space graded by ghost number $\gh$, and $\deg$ is the
grading on $\G$. The space of functions $\Fun(\F)$ becomes
$\ZZ$-graded commutative associative algebra with grading given by
ghost number. The action (\ref{extended BF action}) has ghost number
zero.

\subsubsection*{\bf Remark} Extended $BF$ theory arises naturally in
process of quantizing ordinary $BF$ theory by incorporating ghosts,
ghosts for ghosts etc. and anti-fields for all these (and original
fields).  This set of fields is then organized into a pair of
``superfields'' --- non-homogeneous differential forms $\omega$ and
$p$, and the action may be written in simple form (\ref{extended BF
action}).

\subsection{$P$-structure on space of fields of extended $BF$ theory}
Space $\F$ has a structure of $QP$-manifold. The $P$-structure (odd
simplectic structure) is defined by Batalin-Vilkovisky 2-form (odd
simplectic form) \be\Omega_\BV=<\delta\omega,\delta p>\label{BV
form}\ee so that $\omega$ and $p$ are canonically conjugated w.r.t.
$\Omega_\BV$. The odd simplectic structure on $\F$ induces on space
of functions of fields $\Fun(\F)$ the structure of anti-bracket
algebra (odd Poisson algebra) with usual pointwise associative
(super)commutative product and the anti-bracket, defined as
$$\{f,g\}=f\left(<\frac{\overleftarrow\dd}{\dd
p},\frac{\overrightarrow\dd}{\dd
\omega}>-<\frac{\overleftarrow\dd}{\dd\omega},\frac{\overrightarrow\dd}{\dd
p}>\right)g$$ for a pair of functions $f,g\in \Fun(\F)$. Following
properties hold for anti-bracket:
\begin{itemize}
\item symmetry property $$\{f,g\} = -(-1)^{(\e(f)+1)(\e(g)+1)}\{g,f\}$$
\item Leibniz rule for anti-bracket $$\{f,g\cdot h\} = \{f,g\}\cdot h+(-1)^{(\e(f)+1)\,\e(g)}g\cdot\{f,h\}$$
\item Jacobi identity $$\{\{f,g\},h\}+(-1)^{(\e(h)+1)(\e(f)+\e(g))}\{\{h,f\},g\}+
(-1)^{(\e(f)+1)(\e(g)+\e(h))}\{\{g,h\},f\}
 = 0$$
\end{itemize}
Here $\e$ is notation for parity and we assume that $f,g,h\in
\Fun(\F)$ are functions of definite parity. Batalin-Vilkovisky
Laplacian $\Delta_\BV$ on $\Fun(\F)$ is defined as
$$\Delta_\BV= <\frac{\dd}{\dd\omega},\frac{\dd}{\dd p}>$$
It satisfies
\begin{itemize}
\item the nilpotence property
$$\Delta_\BV^2=0$$
\item the property relating $\Delta_\BV$ and the anti-bracket
$$\Delta_\BV (f\cdot g)=\Delta_\BV f\cdot
g+(-1)^{\e(f)}f\cdot\Delta_\BV \,g+(-1)^{\e(f)}\{f,g\}$$
\end{itemize}

\subsection{$Q$-structure on space of fields of extended $BF$ theory; master equation}
 The $Q$-structure on $\F$ is induced from cohomological
vector field $Q$ on base $\Pi\G$ defined as
$$Q=<d\omega+\frac{1}{2}[\omega,\omega],\frac{\dd}{\dd\omega}>$$
Action (\ref{extended BF action}) on $\F$ is then $S=<p,Q\;\omega>$
and cohomological vector field on $\F$ is a Hamiltonian vector field
generated by $S$: $$Q_\F=\{S,\bullet\}$$ Action (\ref{extended BF
action}) satisfies the quantum master equation \be\Delta_\BV
e^{S/\hbar}=0\label{QME}\ee where
 $\hbar$ is formal infinitesimal
parameter. Since $$\Delta_\BV
e^{S/\hbar}=(\hbar^{-2}\frac{1}{2}\{S,S\}+\hbar^{-1}\Delta_\BV
S)e^{S/\hbar}$$ equation (\ref{QME}) is equivalent to the pair of
equations \be\{S,S\}=0\label{CME}\ee (the classical master equation)
and \be\Delta_\BV S=0\label{QME1}\ee In terms of $Q$, (\ref{CME})
means $Q^2=0$, while (\ref{QME1}) means that
\be\mathrm{div}\,Q=0\label{QME2}\ee This is in turn equivalent to
the fact that volume form $\eta=\prod_\alpha \delta\omega^\alpha$ on
$\Pi\G$ is conserved by vector field $Q$.

\subsection{Extended $BF$ action as generating function of DGLA structure
on $\g\otimes\Omega(M)$}

Action (\ref{extended BF action}) may be viewed as the generating
function for structure constants of differential and Lie bracket on
$\G$. Classical master equation (\ref{CME}) is then equivalent to
the three relations in DGLA $\G$: $d^2=0$, Leibniz identity and
Jacobi identity:
$$\frac{1}{2}\{S,S\}=<p,-d^2\omega>+
<p,-\frac{1}{2}\;d[\omega,\omega]+[d\omega,\omega]>+
\frac{1}{2}<p,[[\omega,\omega],\omega]>=0$$ Vanishing of linear in
$\omega$ term here is equivalent to $d^2=0$, of quadratic term ---
to Leibniz identity, of cubic term --- to Jacobi identity. Equation
(\ref{QME1}) reads
$$\Delta_\BV S=\Str_{\Pi\G}\;[\omega,\bullet]=0$$
This follows from relation $f^b_{ab}=0$ which we demand for
structure constants of gauge algebra $\g$. For example, it holds for
semi-simple gauge algebras.

\subsection{Generalization from extended $BF$ to abstract $BF$ theory}
The construction of extended $BF$ theory admits a natural
generalization to abstract $BF$ theory as follows: take any DGLA
$\G$ instead of differential forms on manifold with values in a
gauge algebra. We should only demand that structure constants of
$\G$ satisfy $f^\beta_{\alpha\beta}=0$. Then construct space of
fields (\ref{space of fields}), which is again a $QP$ manifold. The
action is again given by (\ref{extended BF action}) and it again
satisfies quantum master equation by virtue of general properties of
DGLA: $d^2=0$, Leibniz identity, Jacobi identity, and the property
$f^\beta_{\alpha\beta}=0$ which we demanded for $\G$.

\subsection{Canonical transformations, gauge symmetry, symmetry under diffeomorphisms}
\label{canonical transformations} Infinitesimal canonical
transformation is defined as follows: let $R\in\Fun(\F)$ be some
infinitesimal odd function of fields (the generator of canonical
transformation). Then map
\begin{eqnarray*}
\phi^*_R:\Fun(\F) & \rightarrow & \Fun(\F) \\
 f & \mapsto & f+\{f,R\}
\end{eqnarray*}
is an automorphism of anti-bracket algebra $\Fun(\F)$ (in lowest
order in $R$) due to Leibniz identity and Jacobi identity for the
anti-bracket. Canonical transformation on functions $\phi^*_R$ may
be understood as a pullback of simplectomorphism (in terminology of
Hamiltonian formalism, canonical change of coordinates)
$\phi_R:\F\rightarrow\F$, defined by $\omega\rightarrow
\omega+\{\omega,R\}$ and $p\rightarrow p+\{p,R\}$. Action is
transformed by $\phi^*_R$ as
$$S\mapsto S+\{S,R\}+\hbar\,\Delta_\BV R$$
the last term is due to the fact that action is not a scalar
function but rather a logarithm of density of measure on space of
fields, and thus transforms non-tensorially under change of
coordinates. It may be more transparent to write canonical
transformation of action in terms of exponentials:
$$e^{S/\hbar}\mapsto e^{S/\hbar}+\Delta_\BV (e^{S/\hbar} R)$$
Canonically transformed action leads of course to physically
equivalent theory.

It turns out that two important symmetries of extended $BF$ theory,
the gauge symmetry and symmetry under diffeomorphisms, may be
regarded as special canonical transformations that leave action
invariant. Namely, infinitesimal diffeomorphism, generated by vector
field $v$ on manifold $M$, may be regarded as canonical
transformation with generator $$R_v=<p,\LL_v\omega>$$ where $\LL_v$
is the Lie derivative along $v$, acting on differential forms on
$M$.

The gauge invariance may be formulated in more general setting of
abstract $BF$ theory. The parameter of gauge transformation is
totally even element $\alpha\in [\RR^{1|1}\otimes\G]^{\even}$ (for
extended $BF$ case, this is a totally even $\g$-valued differential
form). The gauge transformation acts on fields by
\begin{eqnarray*}
\omega & \mapsto & \omega+[\omega,\alpha]+d\alpha \\
p & \mapsto & p+[p,\alpha]
\end{eqnarray*}
where $[p,\alpha]$ is the right coadjoint action of $\G$ on $\G^*$.
We see that this gauge transformation is actually a special
canonical transformation with generator
$$R_\alpha=-<p,d\alpha+[\omega,\alpha]>$$
It is instructive to note that $R_\alpha$ may be obtained directly
from the action: \be
R_\alpha=<\alpha,\frac{\dd}{\dd\omega}\,S>\label{generator for gauge
transformation}\ee Gauge invariance of the action follows then from
master equation. This approach allows us to describe gauge
transformation in even more general case of $BF_\infty$ theories,
which we will introduce later.

In case of extended $BF$ theory gauge symmetry we described is
rather a ``super gauge symmetry'', since it mixes components of
$\omega$ and $p$ of different de Rham degree. The ordinary gauge
transformations correspond to the case when gauge parameter $\alpha$
is just a $\g$-valued function: $\alpha\in\G^0$. The form of gauge
transformation of field $\omega$ allows us to call $\omega$ the
"superconnection" in trivial $G$-bundle on $M$. Then
$F=d\omega+\frac{1}{2}[\omega,\omega]$ is naturally the curvature of
superconnection $\omega$.

\section{\bf Effective action for abstract $BF$ theory}
We first describe a general construction of effective action for
abstract $BF$ theory, and then specialize to differential forms
(extended $BF$ case) in the next section.

\subsection{Infrared and ultraviolet fields, chain homotopy,
BV integral for effective action on infrared fields} Let DGLA $\G$
be split into sum of two subcomplexes
$$\G=\G'\oplus\G''$$ with $\G''$ acyclic. We call $\G'$ the infrared
subcomplex and $\G''$ the ultraviolet subcomplex. Names ``infrared''
and ``ultraviolet'' come from physical construction of Wilson
effective action in quantum field theory: one splits fields into
low-frequency (infrared) and high-frequency (ultraviolet) parts and
integrates out the ultraviolet fields to obtain effective action on
infrared fields. Space of fields of $BF$ theory associated to $\G$
is then also split: $\F=\F'\oplus\F''$. Fields
$\omega\in\overline{\Pi\G}$ and $p\in\overline{\G^*}$ are decomposed
into infrared and ultraviolet parts: $\omega=\omega'+\omega''$,
$p=p'+p''$ with $\omega'\in \overline{\Pi\G'}$, $\omega''\in
\overline{\Pi\G''}$, $p'\in\overline{[\G']^*}$,
$p''\in\overline{[\G'']^*}$. Let us denote projectors on $\G'$ and
$\G''$ by $\PP'$ and $\PP''$ respectively.

Let also $K$ be a linear operator on $\G''$ satisfying
\begin{eqnarray}
Kd+dK & = & \PP'' \label{Kd+dK=P''} \\
K^2 & = & 0 \label{K^2=0}
\end{eqnarray}
(a chain homotopy). It is continued to $\G$ by defining
$K|_{\G'}=0$. Starting from chain homotopy $K$ we construct a
Lagrangian submanifold $\LL_K\subset\F''$ as a vector subspace in
$\F''$ defined by equations $K\omega''=0$, $p''K=0$. In other words
$$\LL_K=\Pi\;\mathrm{ker}\,K\oplus [\mathrm{coker}\,K]^*$$
Then we define the effective action $S'$ on $\F'$ via an integral
over $\LL_K$ (``BV integral''): \be
e^{\frac{1}{\hbar}S'(\omega',p';\hbar)}=\frac{1}{N}\int_{\LL_K}e^{\frac{1}{\hbar}
S(\omega'+\omega'',p'+p'')}[\DD\omega''\DD p'']_{\LL_K}\label{BV
integral}\ee Here $[\DD\omega''\DD p'']_{\LL_K}$ is the volume form
on $\LL_K$ and
$$N=\int_{\LL_K}e^{\frac{1}{\hbar}
<p'',d\omega''>}[\DD\omega''\DD p'']_{\LL_K}$$ is the normalization
factor.

\subsection{Perturbative evaluation of BV integral for effective action}
\label{Perturbative evaluation of BV integral for effective action}
We view (\ref{BV integral}) as a formal perturbative definition of
$S'$. If we expand in ultraviolet fields the action in exponent in
integrand of (\ref{BV integral}) we get
\begin{multline}
S(\omega'+\omega'',p'+p'')=S(\omega',p')+<p'',d\omega''>+<p'',\frac{1}{2}[\omega'',\omega'']>+\\
+<p'',\frac{1}{2}[\omega',\omega']>+<p'',[\omega',\omega'']>+<p',[\omega',\omega'']>+
<p',\frac{1}{2}[\omega'',\omega'']>\label{S decomposition}
\end{multline}
The first term here is constant (on $\LL_K$), the second term we
interpret as a free Gaussian action $S''_0=<p'',d\omega''>$. The
other terms are treated as perturbation of $S''_0$ (and hence as
vertices in Feynman rules for (\ref{BV integral})). Let us denote
the normalized expectation value of $f\in \Fun(\LL_K)$ with respect
to $S''_0$ by
$$\ll f\gg_0=\frac{1}{N}\int_{\LL_K}e^{\frac{1}{\hbar}
<p'',d\omega''>}\;f(\omega'',p'')\;[\DD\omega''\DD p'']_{\LL_K}$$
Then propagator $\ll\omega''\otimes p''\gg_0$ viewed as an element
of $\G''\otimes[\G'']^*=\mathrm{End}(\G'')$ is (up to constant
factor) the chain homotopy operator:
$$\ll\omega''\otimes p''\gg_0=-\hbar K$$
This implies that for constant vectors
$\tilde\omega\in\overline{\Pi\G}$, $\tilde{p}\in\overline{\G^*}$ we
have
$$\ll\;\frac{1}{\hbar}<\tilde{p},\omega''>\cdot\frac{1}{\hbar}<p'',\tilde{\omega}>\;\gg_0=
-\frac{1}{\hbar}<\tilde{p},K\tilde\omega>$$ Using this fact and
Wick's theorem we obtain description of values of Feynman graphs for
(\ref{BV integral}) in terms of iterated operation formalism.

First we need to introduce some general notation for a binary
operation iterated on a rooted binary tree. By a rooted binary tree
$T$ we mean an acyclic graph with one vertex of valence 2 (root),
several vertices of valence 1 (leaves) and all other vertices of
valence 3 (internal vertices). This graph comes with fixed planar
structure, i.e. embedding $T\rightarrow \RR^2$ modulo
diffeomorphisms of $\RR^2$, so that each non-leaf vertex has
well-defined left and right children. Let $X$ be a vector
super-space over $\RR$ and $\OO:X\times X\rightarrow X$ a bilinear
map. For a rooted binary tree $T$ with $|T|=n$ leaves, define
$n$-linear map
$$\Iter_{T,\OO}: X^n\rightarrow X$$ by the following iterative
procedure: for $(x_1,\ldots,x_n)$ the $n$-tuple of elements of $X$
decorate each leaf of $T$ with $x_i$ where $i$ is the number of leaf
counted counterclockwise starting from root. Decorate each non-leaf
vertex $v$ with $\OO(x_{v_l},x_{v_r})$ where $x_{v_l}$ and $x_{v_r}$
are elements of $X$ assigned to left and right children of $v$
respectively. We define $\Iter_{T,\OO}(x_1,\ldots,x_n)$ as the value
assigned to root of $T$ by this procedure. We also need the
following modification of this definition: for $\OO,\OO':X\times
X\rightarrow X$ a pair of binary operations we define
$$\Iter_{T,\OO,\OO'}: X^n\rightarrow X$$ by the same procedure as
above with the only difference that in the root we evaluate $\OO'$
on children instead of $\OO$ If we identify binary rooted trees with
binary bracket structures, we have for example
$$\Iter_{((*(**))(**)),\;\OO}\;(x_1,x_2,x_3,x_4,x_5)=\OO(\OO(x_1,\OO(x_2,x_3)),\OO(x_4,x_5))$$
and
$$\Iter_{((*(**))(**)),\;\OO,\OO'}\;(x_1,x_2,x_3,x_4,x_5)=\OO'(\OO(x_1,\OO(x_2,x_3)),\OO(x_4,x_5))$$
Here symbols $*$ denote leaves of the tree.

We also need the general notation for trace operator associated to a
bilinear map and a binary 1-loop graph. Let $L$ be a graph with one
oriented cycle (and fixed planar structure), internal vertices of
valence 3 and $|L|=n$ vertices of valence 1 (leaves). We define
$n$-linear function $\Loop_{L,\OO,X}:X^n\rightarrow\RR$ as follows.
Cutting any edge of the cycle of $L$ produces a tree $T$ with $n+1$
leaves and one leaf marked (it was connected to root before
cutting). We define $\Loop_{L,\OO,X}$ as the super-trace
$$\Loop_{L,\OO,X}(x_1,\cdots,x_n)=\Str_X\Iter_{T,\OO}(x_1,\cdots,x_{i-1},\bt,x_i,\cdots,x_n)$$
where $i$ is the number of the marked leaf (counted counterclockwise
from the root, as before). If we denote 1-loop graphs as trees with
one marked leaf, we have for example
$$\Loop_{(((**)\bt)*),\;\OO,\;X}=\Str_X \OO(\OO(\OO(x_1,x_2),\bt),x_3)$$
Here symbols $*$ denote non-marked leaves and $\bt$ denotes the
marked leaf.

Let $\TTT$ be the set of binary rooted trees and $\LLL$ be the set
of binary 1-loop graphs. Let also $\hat\TTT$ be set of binary rooted
trees without planar structure (i.e. quotient of $\TTT$ over graph
isomorphisms), and $\hat\LLL$ --- the set binary 1-loop graphs
without planar structure.

Having introduced the necessary notation we return to perturbation
theory for (\ref{BV integral}). For every tree $T\in\hat\TTT$ we
define a function $S_T\in \Fun(\F')$ as
\begin{equation}
S_T(\omega',p')
=\frac{1}{|\Aut(T)|}<p',\Iter_{T,\;-K[\bt,\bt],\;[\bt,\bt]}(\omega',\ldots,\omega')>\label{S_T}
\end{equation}
Here  $\Aut(T)$ is the group of automorphisms of $T$ and $|\Aut(T)|$
is its order (factor $\frac{1}{|\Aut(T)|}$ is usually called
``symmetry coefficient'' of the Feynman graph). Expression
(\ref{S_T}) is linear in $p'$ and of degree $|T|$ in $\omega'$. It
does not depend on planar structure on $T$ since binary operations
we are iterating $\OO=-K[\bt,\bt]$ and $\OO'=[\bt,\bt]$ are
commutative on $\overline{\Pi\G'}$, and so the result does not
depend on which child of a vertex we call left and which we call
right.

Analogously, for every binary 1-loop graph $L\in\hat\LLL$ we define
a function $S_L\in \Fun(\Pi \G')$ as the super-trace over $\Pi\G'$:
\begin{equation}
S_L(\omega')
=\frac{1}{|\Aut(L)|}\Loop_{L,\;-K[\bt,\bt],\;\overline{\Pi\G'}}
(\omega',\ldots,\omega')\label{S_L}
\end{equation}
Here $|\Aut(L)|$ is again the order of automorphism group of graph
$L$. Expression (\ref{S_L}) is of degree $|L|$ in $\omega'$. The
independence of (\ref{S_L}) on planar structure on $L$ is checked by
the same argument as for trees.

Now we have all the ingredients to describe the perturbation series
for $S'$ in powers of $\omega'$:
\begin{Theorem}[\bf Perturbation expansion for effective action of abstract $BF$
theory] \label{thm 1} Effective action is linear in $\hbar$:
$$S'(\omega',p';\hbar)=S'^{(0)}(\omega',p')+\hbar S'^{(1)}(\omega')$$
$S'^{(0)}$ is represented as sum over rooted binary trees without
planar structure
\begin{multline}
S'^{(0)}(\omega',p')=S(\omega',p')+\sum_{T\in\hat\TTT:\;|T|\geq
3}S_T(\omega',p')=\\
=S(\omega',p')+\sum_{T\in\hat\TTT:\;|T|\geq
3}\frac{1}{|\Aut(T)|}<p',\Iter_{T,\;-K[\bt,\bt],\;[\bt,\bt]}(\omega',\ldots,\omega')>
\end{multline}
 The first term here is a restriction of $BF$
action in full space $\F$ to subspace $\F'$. $S'^{(1)}$ is a sum
over binary 1-loop graphs $L$ without planar structure
\begin{equation}S'^{(1)}(\omega')=\sum_{L\in\hat{\LLL}}
S_L(\omega')=\sum_{L\in\hat{\LLL}}
\frac{1}{|\Aut(L)|}\Loop_{L,\;-K[\bt,\bt],\;\overline{\Pi\G'}}
(\omega',\ldots,\omega')
\end{equation}
 and does
not depend on $p'$. First terms of perturbation expansions for
$S'^{(0)}$ and $S'^{(1)}$ are:
\begin{multline}S'^{(0)}(\omega',p')=<p',d\omega'>+\frac{1}{2}<p',[\omega',\omega']>-
\frac{1}{2}<p',[K[\omega',\omega'],\omega']>+\\+\frac{1}{2}<p',[K[K[\omega',\omega'],\omega'],\omega']+
\frac{1}{8}<p',[K[\omega',\omega'],K[\omega',\omega']]>+\OO(p'\omega'^5)
\label{S'^0}
\end{multline} and
\begin{multline}S'^{(1)}(\omega')=-\Str\,K[\omega',\bullet]+\frac{1}{2}\Str\,K[K[\omega',\omega'],\bullet]+
\frac{1}{2}\Str\,K[\omega',K[\omega',\bullet]]-\\-\frac{1}{2}\Str\,K[K[K[\omega',\omega'],\omega'],\bullet]-
\frac{1}{2}\Str\,K[K[\omega',\omega'],K[\omega',\bullet]]-
\frac{1}{3}\Str\,K[\omega',K[\omega',K[\omega',\bullet]]]+\OO(\omega'^4)
\label{S'^1}
\end{multline}
\end{Theorem}

\subsection{Properties of effective theory on infrared fields: $QP$-structure on
space of fields, master equation}
Space of infrared fields $\F'=\Pi
T^*(\Pi\G')$ becomes equipped with $QP$-structure in the following
way. The $P$ structure is provided by restriction of BV 2-form on
$\F$ to $\F'$:
$$\Omega'_\BV=\Omega_\BV|_{\F'}=<\delta\omega',\delta p'>$$
Analogously, the BV Laplacian and anti-bracket on $\Fun(\F')$ are
just restrictions of their counterparts on $\Fun(\F)$ to
$\Fun(\F')$. Effective action $S'\in \Fun(\F')$ automatically
satisfies quantum master equation by virtue of general property of
BV integrals:
\begin{multline*}<\frac{\dd}{\dd\omega'},\frac{\dd}{\dd
p'}>e^{\frac{1}{\hbar}S'(\omega',p';\hbar)}=
\frac{1}{N}\int_{\LL_K}(\Delta_\BV-<\frac{\dd}{\dd\omega''},\frac{\dd}{\dd
p''}>)e^{\frac{1}{\hbar} S(\omega,p)}[\DD\omega''\DD
p'']_{\LL_K}=\\
=-\frac{1}{N}\int_{\LL_K}<\frac{\dd}{\dd\omega''},\frac{\dd}{\dd
p''}>e^{\frac{1}{\hbar} S(\omega,p)}[\DD\omega''\DD p'']_{\LL_K}=0
\end{multline*}
In terms of $S'^{(0)}$ and $S'^{(1)}$ the quantum master equation
means \be\{S'^{(0)},S'^{(0)}\}=0\label{CME'}\ee (the classical
master equation for $S'^{(0)}$) and
\be\{S'^{(0)},S'^{(1)}\}+\Delta_\BV S'^{(0)}=0\label{QME'}\ee Hence
tree effective action $S'^{(0)}$ provides a $Q$ structure to $\F'$
--- the cohomological vector field
$$Q_{\F'}=\{S'^{(0)},\bullet\}$$ Since $S'^{(0)}$ is linear in $p'$,
vector field $Q_{\F'}$ is tangent to the base $\Pi\G'$, and defines
on it the cohomological vector field $Q'$ (thus $Q'$ is a
coderivation of $\Fun(\Pi\G')$). In terms of $Q'$ the classical
master equation (\ref{CME'}) is the cohomologicity condition \be
Q'^2=0\label{CME'1}\ee while the equation (\ref{QME'}) means
\be\mathrm{div}\;Q'+Q' S'^{(1)}=0\label{QME'1}\ee

\subsection{Tree effective action on infrared fields as generating function of
$L_\infty$ algebra structure} A well known theorem from \cite{AKSZ}
states that $Q$ - structure on a manifold $\N$ generates an
$L_\infty$ algebra structure on parity-reversed tangent space $\Pi
T_a \N$ in the point $a\in\N$ where $Q$ vanishes. In our case of
effective $BF$ theory we have $\N=\Pi\G'$, $a=0$, due to linearity
of space of fields we identify $\Pi T_0(\Pi\G')$ with $\G'$. Thus
$Q'$ is a generating function for $L_\infty$ algebra structure on
$\G'$.

This is also a special case of Koszul duality: to introduce a
coderivation $Q'$ on commutative associative super algebra of
functions $\Fun(\Pi\G')$ is equivalent to defining $L_\infty$
structure on $\G'$.

\subsection{Construction of $L_\infty$ quasi-isomorphism between
$\G'$ and $\G$ via expectation value map for BV integral;
perturbative series} We can construct an $L_\infty$
quasi-isomorphism $U$ of $L_\infty$ algebra $(\G',Q')$ and DGLA
$(\G,Q)$:
$$U:\Pi\G'\rightarrow\Pi\G$$
Map $U$ is a non-linear deformation of the embedding
$\iota:\G'\rightarrow\G$. The pullback $U^*:\Fun(\Pi\G)\rightarrow
\Fun(\Pi\G')$ is constructed as expectation value map: \be
U^*(f)(\omega')=\frac{\int_{\LL_K}f(\omega)\;e^{\frac{1}{\hbar}
S(\omega,p)}[\DD\omega''\DD
p'']_{\LL_K}}{\int_{\LL_K}e^{\frac{1}{\hbar}
S(\omega,p)}[\DD\omega''\DD p'']_{\LL_K}}\label{U^* as VEV}\ee for
$f\in \Fun(\Pi\G)$. Map $U^*$ can be lifted to pre-$L_\infty$
morphism $U:\Pi\G'\rightarrow\Pi\G$ because $U^*$ is a homomorphism:
$U^*(f g)=U^*(f) U^*(g)$. This is in turn a consequence of the fact
that field $\omega$ is non self-interacting in $BF$ theory. To show
that $U$ is a true $L_\infty$ morphism we need to check that for any
function $f\in \Fun(\Pi\G)$ we have \be Q' U^*(f)=U^*
(Qf)\label{condition for U}\ee This is ensured by the following
argument:
$$\Delta'_\BV (e^{S'/\hbar}U^*(f)) = e^{S'/\hbar}(\frac{1}{\hbar}\{S',U^*(f)\}'+\Delta'_\BV U^*(f))
=\frac{1}{\hbar}e^{S'/\hbar}\{S',U^*(f)\}'=\frac{1}{\hbar}e^{S'/\hbar}
Q'U^*(f)$$ (we put primes here on BV Laplacian and anti-bracket on
$\Fun(\F')$ to distinguish them from their counterparts on full
space $\Fun(\F)$). On the other hand
\begin{multline*}\Delta'_\BV
(e^{S'/\hbar}U^*(f))=\frac{1}{N}\Delta'_\BV\left(\int_{\LL_K}f\;e^{S/\hbar}
[\DD\omega''\DD p'']_{\LL_K}\right)= \\ =
\frac{1}{N}\int_{\LL_K}\Delta_\BV(f\;e^{S/\hbar}) [\DD\omega''\DD
p'']_{\LL_K} =\frac{1}{N}\int_{\LL_K}\frac{1}{\hbar}(Q
f)\;e^{S/\hbar} [\DD\omega''\DD
p'']_{\LL_K}=\frac{1}{\hbar}e^{S'/\hbar} U^*(Q f)
\end{multline*}
Hence (\ref{condition for U}) holds. Fact that $U$ is
quasi-isomorphism is trivial since the embedding $\iota$ obviously
induces an isomorphism of cohomologies $\iota:H^*(\G')\rightarrow
H^*(\G)$. The perturbation expansion for (\ref{U^* as VEV}) gives an
expression for $U$ as sum over binary rooted trees. We summarize
these results in the following statement.
\begin{Theorem}
\label{L_infty morphism} Map $U:\Pi\G'\rightarrow\Pi\G$ defined by
(\ref{U^* as VEV}) is an $L_\infty$ quasi-isomorphism between
$L_\infty$ algebra $(\G',Q')$ and DGLA $(\G,Q)$, and may be expanded
as the following sum over binary rooted trees
\begin{multline}
U(\omega')=\omega'+\sum_{T\in\hat\TTT:\;|T|\geq
2}\frac{1}{|\Aut(T)|}\Iter_{T,\;-K[\bt,\bt]}(\omega',\ldots,\omega')=\\=
\omega'-\frac{1}{2}K[\omega',\omega']+\frac{1}{2}K[K[\omega',\omega'],\omega']-
\frac{1}{2}K[K[K[\omega',\omega'],\omega'],\omega']-
\frac{1}{8}K[K[\omega',\omega'],K[\omega',\omega']]+\cdots
\label{series for U}
\end{multline}
\end{Theorem}

\subsection{1-loop effective action on infrared fields as logarithm of density on $\Pi\G'$}
The 1-loop part of effective action $S'^{(1)}$ has the following
interpretation. Define function $\rho'\in \Fun(\Pi\G')$ as
exponential of $S'^{(1)}$:
$$\rho'(\omega')=e^{S'^{(1)}(\omega')}$$ Then $\rho'$ is a density
on space $\Pi\G'$, such that the volume form
\be\eta'=\rho'\prod_{\alpha'}\delta\omega^{\alpha'} \label{volume
form}\ee on $\Pi\G'$ is conserved by $Q'$ (in the sense that Lie
derivative of $\eta'$ along $Q'$ vanishes). This conservation is
equivalent to (\ref{QME'1}). Another formulation of this
conservation property is hydrodynamical: substituting $S'^{(1)}=\log
\rho'$ into (\ref{QME'1}) we obtain equation
$$\rho'\;\mathrm{div}\;Q'+Q'\,\rho'=0$$
which is known in hydrodynamics as the equation of conservation of
compressible fluid in a stationary flow, with $Q'$ the velocity
field of the flow and $\rho'$ the density of the fluid.

\subsection{Dependence of effective action on chain homotopy}
Our definition of effective $BF$ action (\ref{BV integral}) depends
on choice of chain homotopy operator $K:\G''\rightarrow\G''$. We
will include $K$ as a subscript in notation $S'_K(\omega',p';\hbar)$
while we are interested in $K$-dependence. We formulate a statement
on behaviour of $S'_K$ under infinitesimal changes of chain homotopy
$K\mapsto K+\delta K$. As a consequence of
(\ref{Kd+dK=P''},\ref{K^2=0}), for $K+\delta K$ to be a chain
homotopy (in first order in $\delta K$) the variation $\delta K$
needs to satisfy two properties: $d\,\delta K+\delta K\, d=0$ and
$K\,\delta K+\delta K\, K=0$.
\begin{Theorem}
\label{K-dependence of S'} Effective action $S'_{K+\delta K}$
differs from $S'_K$ by an infinitesimal canonical transformation
\begin{equation}
S'_{K+\delta K}-S'_K=\{S'_K,R_{K,\,\delta K}\}+\hbar\;\Delta_\BV
(R_{K,\,\delta K})
\end{equation}
 and the generator of canonical transformation
$R_{K,\,\delta K}\in \Fun (\F')$ is given by
\begin{equation}R_{K,\,\delta
K}(\omega',p';\hbar)= \left.\frac{\dd}{\dd z}\right|_{z=0}S'_{K+z
K\,\delta K}(\omega',p';\hbar) \label{R from delta K exact}
\end{equation}
where $z\in \RR^{0|1}$ is an odd infinitesimal variable.
Equivalently, the exponential of effective action changes under
$K\mapsto K+\delta K$ by a $\Delta_\BV$-exact term:
\begin{equation}
e^{\frac{1}{\hbar}S'_{K+\delta K}}-e^{\frac{1}{\hbar}S'_{K}}=
\Delta_\BV\left(\hbar\left.\frac{\dd}{\dd z}\right|_{z=0}
e^{\frac{1}{\hbar}S'_{K+zK\,\delta K}}\right)
\end{equation}
\end{Theorem}
For the generator of infinitesimal canonical transformation
$R_{K,\,\delta K}$ we obtain using (\ref{R from delta K exact}) and
series (\ref{S'^0},\ref{S'^1}) the perturbative expansion
\begin{multline}R_{K,\,\delta K}(\omega',p';\hbar)=
\\
= -\frac{1}{2}<p',[K\,\delta K[\omega',\omega'],\omega']>+
\frac{1}{2}<p',[K\,\delta K[K[\omega',\omega'],\omega'],\omega']>-
\\
-\frac{1}{2}<p',[K[K\,\delta K[\omega',\omega'],\omega'],\omega']>+
\frac{1}{4}<p',[K\,\delta K[\omega',\omega'],K[\omega',\omega']]>-\\
-\hbar\;\Str\;K\,\delta K[\omega',\bt] +\hbar\;\Str\;K\,\delta
K[\omega',K[\omega',\bt]]+\OO(p'\omega'^5)+ \OO(\hbar\;\omega'^3)
\end{multline}
This expansion may be interpreted as a sum over binary rooted trees
with one internal edge marked (we put operator $-K\,\delta K$ on the
marked edge and $-K$ on the others, as usual) plus sum over binary
1-loop graphs with one internal edge marked (and the same rule for
assigning operators to edges as for trees).

\subsection{Physical and mathematical interpretations of procedure of inducing
effective action for abstract $BF$ theory} We have two
interpretations of construction for effective action $S'\in
\Fun(\F')$ from abstract $BF$ action $S\in \Fun(\F)$. The first is
the physical interpretation: we construct effective action (in
Wilson sense) for abstract $BF$ theory by integrating out
ultraviolet degrees of freedom (\ref{BV integral}), ending up with
an effective topological theory on the space of infrared fields
$\F'$. The second is mathematical interpretation: starting from DGLA
$\G$ (with additional property $f^\beta_{\alpha\beta}=0$), we
construct $L_\infty$ algebra structure on subcomplex $\G'\subset\G$,
containing all the cohomologies of $\G$: $H^*(\G)\subset\G'$. The
$L_\infty$ operations on $\G'$ are generated by cohomological vector
field $Q'$ on $\Pi\G'$. Additionally we get density function
$\rho'=e^{S'^{(1)}}\in \Fun (\Pi\G')$, defining $Q'$-invariant
measure (\ref{volume form}) on space $\Pi\G'$. We have also built an
$L_\infty$ quasi-isomorphism (\ref{U^* as VEV},\ref{series for U})
between $\G'$ and $\G$.

Apparently, the physically-inspired tool, the BV integral, gives
answers to questions that may be formulated in terms of homotopy
algebra, but are not, to our knowledge, studied. Especially, not
only we have the fact of existence of quasi-isomorphism between $\G$
and $\G'$, but we have expression for it in terms of BV integral.
Further, the $Q'$-invariant measure $\rho'$ is a new object for
homotopy algebra. The pair $(Q',\rho')$ of a cohomological vector
field on $\Pi\G'$ and $Q'$-invariant measure on $\Pi\G'$ should be
considered as defining a structure of ``quantum $L_\infty$ algebra''
on $\G'$.

\subsection{Generalization to $BF_\infty$ theories. Class of $BF_\infty$ theories as ``closure'' of
class of abstract $BF$ theories with respect to procedure of
inducing effective action} Effective theory for abstract $BF$ theory
belongs to a wider class of $BF$ theories, which we call
$BF_\infty$. We define a $BF_\infty$ theory in the following way:
let $(\G,Q,\rho)$ be any $L_\infty$ algebra with $\rho\in
\Fun(\Pi\G)$ a $Q$-invariant density on $\Pi\G$. Then the space of
fields is (as for extended $BF$ and abstract $BF$ case)
$$\F=\Pi T^*(\Pi\G)$$ and the action $S\in \Fun(\F)$
\be
S(\omega,p;\hbar)=<p,Q\,\omega>+\hbar\;\log\rho(\omega)\label{BF_infty
action via Q,rho}\ee We keep the notation $S^{(0)}$ for
$<p,Q\,\omega>$ and $S^{(1)}$ for $\log\rho$. The $BF_\infty$ action
automatically satisfies quantum master equation. This action is also
invariant under gauge transformations --- canonical transformations
on $\F$ with generator
$$R_\alpha=<\alpha,\frac{\dd}{\dd \omega}\,S^{(0)}>$$ where gauge
parameter $\alpha$ belongs to $[\RR^{1|1}\otimes\G]^\even$.
Invariance of action under gauge transformation follows directly
from master equation. This argument is a straightforward
generalization of argument from section \ref{canonical
transformations}.

If $\G$ is split into a sum of two subcomplexes $\G=\G'\oplus\G''$
with $\G''$ acyclic, and $K:\G''\rightarrow\G''$ is the chain
homotopy, we can use BV integral (\ref{BV integral}) to define
effective action $S'\in \Fun(\F')$ on $\F'=\Pi T^*(\Pi\G')$. Then
the effective theory on $\F'$ is again $BF_\infty$ theory. Class of
$BF_\infty$ theories may be regarded as the closure of class of
abstract $BF$ theories with respect to operation of inducing
effective action.

\subsection{Perturbative expansion for effective action of
$BF_\infty$ theory}\label{effective action for BF_infty} There are
now more admissible Feynman graphs in perturbation expansion for
$S'$ then for case of inducing from abstract $BF$ theory (subsection
\ref{Perturbative evaluation of BV integral for effective action}),
due to the fact that action $S(\omega'+\omega'',p'+p'';\hbar)$ now
contains vertices of order $\OO(p''\omega''^3)$,$\OO(p''\omega''^4)$
etc. as well as vertices of order $\OO(\hbar\omega'')$,
$\OO(\hbar\omega''^2)$ etc.

Let us introduce the the obvious generalization of $\Iter$ and
$\Loop$ for case of rooted trees and 1-loop graphs without
restriction on vertices to be trivalent (non-binary case). Let $X$
be a vector super-space over $\RR$ and $\{\OO_k\}_{k\geq
2}=\{\OO_2,\OO_3,\ldots\}$ be a collection of polylinear maps
$\OO_k: X^k\rightarrow X$. Let $T$ be a rooted tree with $|T|=n$
vertices of valence 1 (leaves), one root of valence $\geq 2$ and all
other vertices (internal vertices) of valence $\geq 3$ (we also mean
that $T$ comes with planar structure). We define the $n$-linear map
$\Iter_{T,\{\OO_k\}}:X^n\rightarrow X$ by the same iterative
procedure as for binary trees, with only difference that we decorate
each vertex (internal or root) with $k$ children by $\OO_k$
evaluated on values assigned to children. Map
$\Iter_{T,\{\OO_k\},\{\OO'_k\}}:X^n\rightarrow X$ is defined
analogously where in the root we evaluate operator $\OO'_k$ instead
of $\OO_k$, where $k$ is the valence of root. For example,
$$\Iter_{((*(***))*(**)),\;\{\OO_k\},\;\{\OO'_k\}}(x_1,x_2,x_3,x_4,x_5,x_6,x_7)=
\OO'_3(\OO_2(x_1,\OO_3(x_2,x_3,x_4)),x_5,\OO_2(x_6,x_7))$$ We also
need a special case when operators $\OO'_k$ take values in $\RR$
instead of $X$. The definition of $\Iter$ does not change and it
becomes an $\RR$-valued map
$\Iter_{T,\{\OO_k\},\{\OO'_k\}}:X^n\rightarrow \RR$. If we include
the unary operator $\OO'_1$ in the list of operators $\{\OO'_k\}$
then we mean that trees with univalent root are allowed for this
case.

Let $L$ be a 1-loop graph: a graph with one oriented cycle, $|L|=n$
vertices of valence 1 --- leaves, and with all other vertices
(internal ones) of valence $\geq 3$. We define
$\Loop_{L,\;\{\OO_k\},\;X}: X^n\rightarrow\RR$ in complete analogy
with binary case: we cut the cycle to transform $L$ into a rooted
tree $T$ with one marked leaf and set
$$\Loop_{L,\{\OO_k\},X}(x_1,\ldots,x_n)=\Str_X\Iter_{T,\{\OO_k\}}(x_1,\ldots,x_{i-1},\bt,x_i,\ldots,x_n)$$
where $i$ is the number of marked leaf. For example,
$$\Loop_{((**)(*\bt *)),\;\{\OO_k\},\;X}(x_1,x_2,x_3,x_4)=\Str_X\;\OO_2(\OO_2(x_1,x_2),\OO_3(x_3,\bt,x_4))$$

We also introduce notation $\TTT_\infty$, $\LLL_\infty$ for the sets
of rooted trees and 1-loop graphs with planar structure, and
notation $\hat\TTT_\infty$, $\hat\LLL_\infty$ for the corresponding
sets factorized over graph isomorphisms (i.e. with planar structure
forgotten).

Now we return to description of perturbation series for effective
action of $BF_\infty$ theory. Let Taylor series for $Q$ be
$$Q=<\sum_{n=1}^\infty \frac{1}{n!}\;l^{(n)}(\omega,\ldots,\omega),\frac{\dd}{\dd\omega}>$$
with $l^{(n)}:(\Pi\G)^{\otimes n}\rightarrow\G$ the set of
super-antisymmetric polylinear maps (the $L_\infty$ algebra
operations on $\G$) and Taylor series for $S^{(1)}$ be
$$S^{(1)}=\sum_{n=1}^\infty \frac{1}{n!}\;q^{(n)}(\omega,\ldots,\omega)$$
with $q^{(n)}\in \Fun (\Pi\G)$ the set of super-antisymmetric
polylinear functions on $\Pi\G$. Let us formulate the generalization
of theorem \ref{thm 1} for $BF_\infty$ case.
\begin{Theorem}
Effective action of $BF_\infty$ theory has the form
$$S'(\omega',p';\omega)=S'^{(0)}(\omega',p')+\hbar\; S'^{(1)}(\omega')$$
with $S'^{(0)}$ expanded as a sum over rooted trees without planar
structure as
\begin{multline}
S'^{(0)}(\omega',p')=\\=<p',l^{(1)}(\omega')>+\sum_{T\in\hat\TTT_\infty}\frac{1}{|\Aut(T)|}
<p',\Iter_{T,\;\{-K\circ l^{(k)}\}_{k\geq 2},\;\{l^{(k)}\}_{k\geq
2}}(\omega',\ldots\omega')> \label{S'^0 for BF_infty exact}
\end{multline}
and $S'^{(1)}$ is expanded as
\begin{multline}
S'^{(1)}(\omega')=\sum_{L\in\hat\LLL_\infty}\frac{1}{|\Aut(L)|}
\Loop_{L,\;\{-K\circ l^{(k)}\}_{k\geq
2},\;\overline{\Pi\G'}}(\omega',\ldots,\omega')+\\+\sum_{T\in\hat\TTT_\infty}
\frac{1}{|\Aut(T)|}\Iter_{T,\{\-K\circ l^{(k)}\}_{k\geq
2},\{q^{(k)}\}_{k\geq 1}}(\omega',\ldots,\omega') \label{S'^1 for
BF_infty exact}
\end{multline}
First terms in (\ref{S'^0 for BF_infty exact}) are given by
\begin{multline}S'^{(0)}(\omega',p')=<p',l^{(1)}(\omega')>+\frac{1}{2}<p',l^{(2)}(\omega',\omega')>+
\frac{1}{6}<p',l^{(3)}(\omega',\omega',\omega')>-\\-\frac{1}{2}<p',l^{(2)}(Kl^{(2)}(\omega',\omega'),\omega')>+
\frac{1}{24}<p',l^{(4)}(\omega',\omega',\omega',\omega')>-\\-
\frac{1}{6}<p',l^{(2)}(Kl^{(3)}(\omega',\omega',\omega'),\omega')>-
\frac{1}{4}<p',l^{(3)}(Kl^{(2)}(\omega',\omega'),\omega',\omega')>+\\
+\frac{1}{2}<p',l^{(2)}(Kl^{(2)}(Kl^{(2)}(\omega',\omega'),\omega'),\omega')>+
\frac{1}{8}<p',l^{(2)}(Kl^{(2)}(\omega',\omega'),Kl^{(2)}(\omega',\omega'))>+\OO(p'\omega'^5)
\label{S'^0 for BF_infty}
\end{multline}
and the first terms in (\ref{S'^1 for BF_infty exact}) are
\begin{multline}
S^{(1)}(\omega')=q^{(1)}(\omega')-\Str\;Kl^{(2)}(\omega',\bullet)+\frac{1}{2}
q^{(2)}(\omega',\omega')-\frac{1}{2}q^{(1)}(Kl^{(2)}(\omega',\omega'))+\\
+\frac{1}{2}\Str\;Kl^{(2)}(Kl^{(2)}(\omega',\omega'),\bullet)+
\frac{1}{2}\Str\;Kl^{(2)}(\omega',Kl^{(2)}(\omega',\bullet))+
\frac{1}{6}q^{(3)}(\omega',\omega',\omega')-\frac{1}{2}q^{(2)}(Kl^{(2)}(\omega',\omega'),\omega')-\\-
\frac{1}{6}q^{(1)}(Kl^{(3)}(\omega',\omega',\omega'))+
\frac{1}{2}q^{(1)}(Kl^{(2)}(Kl^{(2)}(\omega',\omega'),\omega'))
+\frac{1}{6}\Str\;Kl^{(2)}(Kl^{(3)}(\omega',\omega',\omega')),\bullet)-\\-
\frac{1}{2}\Str\;Kl^{(2)}(Kl^{(2)}(Kl^{(2)}(\omega',\omega'),\omega'),\bullet)-
\frac{1}{2}\Str\;Kl^{(2)}(Kl^{(2)}(\omega',\omega'),Kl^{(2)}(\omega',\bullet))-\\-
\frac{1}{3}\Str\;Kl^{(2)}(\omega',Kl^{(2)}(\omega',Kl^{(2)}(\omega',\bullet)))+\OO(\omega'^4)
\label{S'^1 for BF_infty}
\end{multline}
with the super traces taken in $\overline{\Pi\G'}$.
\end{Theorem}

\subsection{Effective action on $\Pi T^*(\Pi H^*(\G))$ as iterative limit.
Case of limiting effective action for extended $BF$ theory, Massey
operations on cohomologies} The procedure of inducing effective
action, starting from $BF_\infty$ theory built on $L_\infty$ algebra
$\G$ can be iterated, and reaches the iterative limit on subcomplex
$\G'=H^*(\G)$ consisting of cohomologies of $\G$. Tree part of the
corresponding effective action generates $L_\infty$ algebra
structure on cohomologies $H^*(\G)$. In particular, when we start
from extended $BF$ theory on manifold $M$, so that
$\G=\g\otimes\Omega(M)$, the iterative limit of inducing effective
action is reached on de Rham cohomologies of $M$ with values in
$\g$: $\G'=\g\otimes H^*_{dR}(M)$. The induced $L_\infty$ algebra
structure on $\g\otimes H^*_{dR}(M)$ generates Massey operations on
de Rham cohomologies $H^*_{dR}(M)$. The 1-loop part of effective
action $S'^{(1)}\in \Fun(\g\otimes\Pi H^*_{dR}(M))$ should then be
interpreted as a generating function for ``quantum Massey
operations'' on $H^*_{dR}(M)$.

\subsection{Iterated induction as parallel transport in category of retracts}
\label{category of retracts} We now proceed to more formal
description of iterated induction. Let $\G$ be a cochain complex,
and suppose that $BF_\infty$ theory on $\G$ (that is, with space of
fields $\Pi T^*(\Pi\G)$) is defined by (\ref{BF_infty action via
Q,rho}) by a pair $(Q,\rho)$ --- a cohomological vector field on
$\Pi\G$ and $Q$-invariant measure on $\Pi\G$. Let $\Ret_\G$ be the
category of retracts of $\G$. Its objects are subcomplexes
$\G'\subset \G$, containing all cohomology of $\G$. Objects
constitute a partially ordered set w.r.t. inclusion: if
$\G',\G''\in\Ret_\G$ and $\G''\subset\G'$, we say that $\G'$ is
larger then $\G''$. Category $\Ret_\G$ possesses the largest object
--- full complex $\G$, and set of smallest objects, corresponding to
different embeddings of cohomologies $H^*(\G)$ into $\G$. Morphisms
in $\Ret_\G$ are retractions: for $\G''\subset\G'$ a pair of objects
(subcomplexes), $\PP:\G'\rightarrow\G''$ a projection and
$K:\ker\PP\rightarrow\ker\PP$ a chain homotopy operator, contracting
$\G'$ onto $\G''$, we associate to the pair $(\PP,K)$ a morphism
$m_{\PP,K}:\G'\rightarrow\G''$. Thus morphisms are always from
larger object to smaller one, and for such a pair of objects there
are typically many morphisms. There are no nontrivial automorphism
in $\Ret_\G$: the only automorphism for each object $\G'$ is the
identity.

Now $BF_\infty$ theory on any object $\G'\in\Ret_\G$ is defined by a
pair $(Q,\rho)_{\G'}$ --- a cohomological vector field and
$Q$-invariant measure on $\Pi\G'$. If $m_{\PP,K}$ is a morphism from
(larger object) $\G'$ to (smaller object) $\G''$, the operation of
induction of $BF_\infty$ theory from $\G'$ to $\G''$, using
projection $\PP$ to define separation of fields into infrared and
ultraviolet parts, and using chain homotopy operator $K$ to define
Lagrangian manifold $\LL_K$ for BV integral, may be viewed as
``parallel transport'' of $(Q,\rho)$ structure from $\G'$ to $\G''$
along morphism $m_{\PP,K}$:
$$\I_{\PP,K}: (Q,\rho)_{\G'}\mapsto(Q,\rho)_{\G''}$$ where $\I_{\PP,K}$
denotes induction. Iterated induction is then interpreted as the
parallel transport along a chain of morphisms. This parallel
transport also respects composition of morphisms: if
$m_{\PP_2,K_2}\circ m_{\PP_1,K_1}=m_{\PP_3,K_3}$ then
$\I_{\PP_2,K_2}\circ\I_{\PP_1,K_1}=\I_{\PP_3,K_3}$. In particular
this means that iterated induction can always be reduced to
induction in one move.

Another equivalent picture may be useful. Category $\Ret_\G$
contains isomorphic objects that are different embeddings of the
same complex into $\G$. We may factorize $\Ret_\G$ over chain
isomorphisms. We denote the factorized category $\Ret^\circ_\G$ (one
might call it "category of abstract retracts"). Its objects are
abstract chain complexes $\G'$ that can be embedded into $\G$ and
with cohomologies coinciding with cohomologies of $\G$. This
category has only one smallest object
--- the complex of cohomologies $H^*(\G)$, and one largest object --- whole
$\G$. A morphism $m_{\iota,\PP,K}:\G'\rightarrow\G''$ is now
specified by a triple $(\iota,\PP,K)$ where $\iota$ is the embedding
(injective chain map) $\iota:\G''\rightarrow\G'$, $\PP$ is
projection $\PP:\G'\rightarrow\G''$ (surjective chain map satisfying
$\PP\circ\iota=\id_{\G''}$) and $K:\ker\PP\rightarrow\ker\PP$ is the
chain homotopy, contracting $G'$ onto $\G''$. This category
$\Ret^\circ_\G$ has fewer objects, but more morphisms between two
given objects than in $\Ret_\G$. In particular, there are nontrivial
automorphisms for objects of $\Ret^\circ_\G$, which correspond to
chain automorphisms of complexes. We may understand operation of
inducing $BF_\infty$ theory as parallel transport of $(Q,\rho)$
structure on objects of $\Ret^\circ_\G$ along morphisms in complete
analogy with $\Ret_\G$.

Interpretation of induction of $BF_\infty$ theory in terms of
category of retracts allows us to understand Wilson-type
renormalization of simplicial $BF$ theory under aggregation of
triangulation in terms of holonomy of the parallel transport $\I$.

\subsection{Towards state-sum for $BF_\infty$ theory}
\label{state-sum def} The state-sum $Z(\G)$ for $BF_\infty$ theory
on $\Pi T^*(\Pi\G)$ may be defined as follows: induce effective
action $S'$ on $\F'=\Pi T^*(\Pi H^*(\G))$, then integrate the
exponential of effective action along the base of $\F'$ (this is our
choice of Lagrangian submanifold in $\F'$): \be Z(\G)=\int_{\Pi
H^*(\G)} e^{S'/\hbar} \DD\omega'= \int_{\Pi
H^*(\G)}\rho'\DD\omega'\label{state-sum}\ee with
$\rho'=e^{S'^{(1)}}$ the induced density function on $\Pi H^*(\G)$.
The integral (\ref{state-sum}) over whole space $\Pi H^*(\G)$ can
diverge, and there should exist some ``non-perturbative'' reason,
why we should regularize this integral. One possible regularization
is to integrate over some domain in $\Pi H^*(\G)$, for instance over
connected component of support of $\rho'$, containing zero. However,
we do not have a good explanation, why one should use this
regularization for state-sum.

\section{\bf Effective action for extended $BF$ theory on a triangulation}
\label{Effective action for extended BF theory on a triangulation}
We now proceed to specializing the construction of effective action
to the case of constructing effective action of extended $BF$ theory
on a triangulated manifold.

\subsection{Whitney complex of a simplicial complex}
Let us recall the concept of Whitney complex of a simplicial complex
(see \cite{Getzler}). Let $\Delta^n$ be a standard geometrical
$n$-simplex with barycentric coordinates $t_0,\ldots,t_n$ subject to
relation $t_0+\cdots+t_n=1$ and inequalities $t_0\geq
0,\cdots,t_n\geq 0$. We introduce a set of special piecewise-linear
differential forms on $\Delta^n$: \be\chi_{i_0\cdots \;i_k}=
k!\sum_{r=0}^k (-1)^r t_{i_r}
dt_{i_0}\wedge\cdots\widehat{dt}_{i_r}\cdots\wedge
dt_{i_k}\label{chi}\ee where hat means exclusion. Forms
$\chi_\sigma$ are associated to subsets
$\{i_0,\ldots,i_k\}\subset\{0,\ldots,n\}$ or, equivalently, to faces
of $\Delta^n$. Following properties hold for forms $\chi_\sigma$:
\begin{itemize}
\item for $\sigma,\sigma'$ faces of $\Delta^n$
$$\int_{\sigma'}\chi_\sigma=\left\{\begin{array}{ll} 1 &
\mbox{if $\sigma=\sigma'$} \\ 0 & \mbox{otherwise}
\end{array}\right.$$
\item de Rham differential acts on forms $\chi_\sigma$ as
$$d\chi_{i_0\ldots \;i_k}=\sum_{j=0}^n\chi_{j i_0\cdots \;i_k}$$
\end{itemize}

Linear space spanned by forms $\chi_\sigma$ is closed under de Rham
differential and is called Whitney complex of $\Delta^n$. We denote
it $\Omega_W(\Delta^n)$. Elements of $\Omega_W(\Delta^n)$ (linear
combinations of forms $\chi_\sigma$) are called Whitney forms. There
is a natural isomorphism between Whitney complex
$\Omega_W(\Delta^n)$ and complex $C^*(\Delta^n)$ of simplicial
cochains on $\Delta^n$ that identifies basis cochains $e_\sigma$
with forms $\chi_\sigma$. The de Rham differential is identified
with the coboundary operator on $C^*(\Delta^n)$. Canonical pairing
between chains and cochains on $\Delta^n$ is interpreted as integral
of Whitney form over a chain.

Let now $\Xi$ be a simplicial complex. The Whitney complex
$\Omega_W(\Xi)$ on $\Xi$ is glued from Whitney complexes on
simplices of $\Xi$ with $\Omega_W(\sigma)|_{\sigma\cap\sigma'}$ and
$\Omega_W(\sigma')|_{\sigma\cap\sigma'}$ identified. The cocycle
condition for this gluing is ensured by ``compatibility'' of Whitney
complexes on a simplex $\sigma$ and its face $\sigma'\subset\sigma$:
$$\Omega_W(\sigma)|_{\sigma'}=\Omega_W(\sigma')$$
 A Whitney form $\alpha\in\Omega_W(\Xi)$ is a differential
form on $\Xi$ such that its restrictions to all simplices of $\Xi$
are Whitney forms: $\alpha|_\sigma\in\Omega_W(\sigma)$. Basis forms
$\chi_\sigma$ are associated to each simplex $\sigma\in\Xi$. Form
$\chi_\sigma$ is defined by (\ref{chi}) on each simplex
$\sigma'\in\Xi$ containing $\sigma$ as a face, and by zero on all
other simplices. Whitney complex $\Omega_W(\Xi)$ can again be
identified with complex of simplicial cochains $C^*(\Xi)$, as in the
case of one simplex $\Xi=\Delta^n$. By this identification we chose
special representatives for simplicial cochains in de Rham algebra
$\Omega(\Xi)$ --- the Whitney forms.

Projection $\PP_W:\Omega(\Xi)\rightarrow\Omega_W(\Xi)$ is defined as
$$\PP_W(\alpha)=\sum_{\sigma\in\Xi}\left(\int_\sigma\alpha\right)\chi_\sigma$$

\subsection{Chain homotopy between $\Omega(\Xi)$ and $\Omega_W(\Xi)$: Dupont's construction}
\label{Dupont's construction} Consider first the case of one simplex
$\Xi=\Delta^n$. We cite the Dupont's construction of chain homotopy
between $\Omega(\Delta^n)$ and $\Omega_W(\Delta^n)$ from
\cite{Getzler}, adjusting it to our notations.

Given a vertex $[i]$ of the $n$-simplex $\Delta^n$, define the
dilation map
$$\phi_i:[0,1]\times\Delta^n\rightarrow\Delta^n$$
by the formula
$$\phi_i(u,t_0,\ldots,t_n)=(ut_0,\ldots,ut_i+(1-u),\ldots,ut_n)$$
Let $\pi:[0,1]\times\Delta^n\rightarrow\Delta^n$ be the projection
on the second factor, and let
$\pi_*:\Omega^*([0,1]\times\Delta^n)\rightarrow\Omega^{*-1}(\Delta^n)$
be integration over the first factor. Define operators
$$h^i:\Omega^*(\Delta^n)\rightarrow\Omega^{*-1}(\Delta^n)$$
by the formula
$$h^i\alpha=\pi_*\phi^*_i\alpha$$
Let $\ev^i:\Omega(\Delta^n\rightarrow\RR)$ be evaluation at vertex
$[i]$. Stokes's theorem implies that $h^i$ is the chain homotopy
between the identity and $\ev^i$:
$$dh^i+h^i d=\id-\ev^i$$
Operators $h^i$ also satisfy $$h^i h^j+h^j h^i=0$$ The operator \be
K_{\Delta^n}=\sum_{k=0}^{n-1}(-1)^k\sum_{0\leq i_0<\cdots<i_k\leq
n}\chi_{i_0\ldots\;i_k}h^{i_k}\cdots h^{i_0} \label{Dupont's K}\ee
was introduced by Dupont. Dupont proved the following explicit form
of de Rham theorem: \be
dK_{\Delta^n}+K_{\Delta^n}d=\id-\PP_W\label{chain homotopy on
simplex}\ee Thus $K_{\Delta^n}$ is a chain homotopy between
$\id:\Omega(\Delta^n)\rightarrow\Omega(\Delta^n)$ and $\PP_W$. The
following compatibility property holds: if $\sigma$ is a face of
$\Delta^n$ and $\alpha\in\Omega(\Delta^n)$ then
\be(K_{\Delta^n}\alpha)|_\sigma=K_\sigma(\alpha|_\sigma)\label{compatibility
for K}\ee

Now let $\Xi$ be any simplicial complex. We then define
$K_\Xi:\Omega^*(\Xi)\rightarrow\Omega^{*-1}(\Xi)$ by
\be(K_\Xi\alpha)|_\sigma=K_\sigma (\alpha|_\sigma)\label{K on
simplicial complex}\ee for any simplex $\sigma\in\Xi$. This
definition is self-consistent due to (\ref{compatibility for K}).
Operator $K_\Xi$ is a chain homotopy between identity
$\id:\Omega(\Xi)\rightarrow\Omega(\Xi)$ and projection
$\PP_W:\Omega(\Xi)\rightarrow\Omega_W(\Xi)$ \be dK_\Xi+K_\Xi
d=\id-\PP_W\label{chain homotopy on simplicial complex}\ee

\subsection{Effective action of extended $BF$ theory on triangulation:
factorization of BV integral, reducing the problem to single
simplex} Let $M$ be a $D$-dimensional manifold with corners, and let
$\Xi$ be some triangulation of $M$. We split de Rham algebra
$\Omega(M)$ into sum of two subcomplexes:
$$\Omega(M)=\Omega_W(\Xi)\oplus\Omega''(\Xi)$$
with Whitney complex playing the role of infrared subcomplex,
$\Omega''(\Xi)$ the ultraviolet subcomplex. The latter consists of
differential forms $\alpha''$ such that $\int_\sigma\alpha''=0$ for
any simplex $\sigma\in\Xi$. The space of fields of extended $BF$
theory $\F=\Pi T^*(\Pi(g\otimes\Omega(M)))$ is then split into space
of infrared fields $\F'=\Pi T^*(\Pi(g\otimes\Omega_W(\Xi)))$ and
space of ultraviolet fields $\F''=\Pi
T^*(\Pi(g\otimes\Omega''(\Xi)))$. We use BV integral (\ref{BV
integral}) to define effective action $S_\Xi$ on $\F'$. The
Lagrangian manifold over which we integrate in (\ref{BV integral})
is defined by Dupont's chain homotopy operator $K_\Xi$.

Let us split the space of ultraviolet forms into subspaces
enumerated by simplices of $\Xi$:
$$\Omega''(\Xi)=\bigoplus_{\sigma\in\Xi}\Omega''(M,\sigma)$$
where $\Omega''(M,\sigma)$ is the space of forms supported on the
interior of $\sigma$ (and vanishing on its boundary), with zero
integral over $\sigma$:
$$\Omega''(M,\sigma)=\{\alpha''_\sigma\in\Omega(M):\;\alpha''_\sigma|_{M\backslash\sigma}=0,\;
\alpha''_\sigma|_{\dd\sigma}=0,\int_\sigma\alpha''_\sigma=0\}$$
Field $\omega\in\overline{\Pi(\g\otimes\Omega(M))}$ is then
decomposed as \be\omega=\sum_{\sigma\in\Xi}\omega^\sigma
\chi_\sigma+
\sum_{\sigma\in\Xi}\omega''_{(\sigma)}=\sum_{\sigma\in\Xi}\omega'_{(\sigma)}
+ \sum_{\sigma\in\Xi}\omega''_{(\sigma)}\label{omega
decomposition}\ee where $\omega^\sigma\in\Pi\g$ if $\sigma$ is
even-dimensional and $\omega^\sigma\in\g$ if $\sigma$ is
odd-dimensional;
$\omega''_{(\sigma)}\in\overline{\Pi(\g\otimes\Omega''(M,\sigma))}$.
Field $p\in [\Omega(M)]^*\otimes g^*$ is decomposed correspondingly:
\be p=\sum_{\sigma\in\Xi}e^\sigma
p_\sigma+\sum_{\sigma\in\Xi}p''_{(\sigma)}=\sum_{\sigma\in\Xi}p'_{(\sigma)}+\sum_{\sigma\in\Xi}p''_{(\sigma)}\label{p
decomposition}\ee where $p_\sigma\in\g$ if $\sigma$ is
even-dimensional, $p_\sigma\in\Pi\g$ if $\sigma$ is odd-dimensional,
$e^\sigma$ are basis simplicial chains on $\Xi$,
$p''_{(\sigma)}\in\overline{[\Omega(M,\sigma)]^*\otimes\g^*}$. We
use here the identification of Whitney coforms on $\Xi$ and
simplicial chains: $[\Omega_W(\Xi)]^*=C_*(\Xi)$. Substituting
decompositions (\ref{omega decomposition},\ref{p decomposition})
into extended $BF$ action (\ref{extended BF action}), we get
(omitting terms with vanishing support)
\begin{multline}
S(\omega,p)=<p,d\omega+\frac{1}{2}[\omega,\omega]>=\\
=\sum_{\sigma\in\Xi}\left(<p'_{(\sigma)},\sum_{\sigma_1\subset\sigma}d\omega'_{(\sigma_1)}+\frac{1}{2}
\sum_{\sigma_1,\sigma_2\subset\sigma}[\omega'_{(\sigma_1)},\omega'_{(\sigma_2)}]+
\sum_{\sigma_1\subset\sigma}[\omega'_{(\sigma_1)},\omega''_{(\sigma)}]+\frac{1}{2}
[\omega''_{(\sigma)},\omega''_{(\sigma)}]>+\right.\\
\left.+<p''_{(\sigma)},d\omega''_{(\sigma)}+\frac{1}{2}
\sum_{\sigma_1,\sigma_2\subset\sigma}[\omega'_{(\sigma_1)},\omega'_{(\sigma_2)}]+
\sum_{\sigma_1\subset\sigma}[\omega'_{(\sigma_1)},\omega''_{(\sigma)}]+\frac{1}{2}
[\omega''_{(\sigma)},\omega''_{(\sigma)}]>\right)=\\
=\sum_{\sigma\in\Xi}S\left(\sum_{\sigma_1\subset\sigma}\omega'_{(\sigma_1)}+\omega''_{(\sigma)},
p'_{(\sigma)}+p''_{(\sigma)}\right)
\end{multline}
Hence the BV integral (\ref{BV integral}) factorizes:
\begin{equation}
\int_{\LL_{K_\Xi}} e^{\frac{1}{\hbar}S(\omega,p)} [\DD\omega''\DD
p'']_{\LL_{K_\Xi}}=\prod_{\sigma\in\Xi}\int_{\LL_{K_\sigma}}
e^{\frac{1}{\hbar}S(\omega'|_\sigma+\omega''_{(\sigma)},p'_{(\sigma)}+p''_{(\sigma)})}[\DD\omega''_{(\sigma)}\DD
p''_{(\sigma)}]_{\LL_{K_\sigma}}\label{factorization of BV integral
on simplicial complex}
\end{equation}
Here we use that
$\omega'|_\sigma=\sum_{\sigma_1\subset\sigma}\omega'_{(\sigma_1)}$.
The factorization of measure in (\ref{factorization of BV integral
on simplicial complex}) is due to ``simplicial locality'' of $K_\Xi$
(\ref{K on simplicial complex}). It follows that the effective
action $S_\Xi$ on infrared fields splits into sum of contributions
of individual simplices of $\Xi$.
\begin{Theorem}[\bf Separation of variables for $S_\Xi$]
\label{factorization thm} Effective action $S_\Xi$ on $\F'=\Pi
T^*(\Pi(\g\otimes\Omega_W(\Xi)))$ splits as \be
S_\Xi(\omega',p';\hbar)=\sum_{\sigma\in\Xi}\bar{S}_\sigma(\omega'|_\sigma,p'_{(\sigma)};\hbar)\label{reduction
from simplicial complex to simplex}\ee where functions
$\bar{S}_\sigma$ are defined by following ``elementary'' BV
integrals: \be
e^{\frac{1}{\hbar}\bar{S}_\sigma(\omega'|_\sigma,p'_{(\sigma)};\hbar)}=
\frac{\int_{\LL_{K_\sigma}}
e^{\frac{1}{\hbar}S(\omega'|_\sigma+\omega''_{(\sigma)},p'_{(\sigma)}+p''_{(\sigma)})}[\DD\omega''_{(\sigma)}\DD
p''_{(\sigma)}]_{\LL_{K_\sigma}}}{\int_{\LL_{K_\sigma}}e^{\frac{1}{\hbar}<p''_{(\sigma)},d\omega''_{(\sigma)}>}
[\DD\omega''_{(\sigma)}\DD p''_{(\sigma)}]_{\LL_{K_\sigma}}}
\label{elementary BV integral on simplex}\ee
\end{Theorem}
Thus the task of calculating effective action of extended $BF$
theory on any triangulated manifold $(M,\Xi)$ is reduced to the
series of universal problems: calculate (\ref{elementary BV integral
on simplex}) for a simplex of each dimension $\sigma=\Delta^n$ with
$n=0,1,2,\ldots$

Notice that the elementary BV integral (\ref{elementary BV integral
on simplex}) does not define an effective action of $BF$-type
theory, since $\bar{S}_\sigma$ is a function on space
$$\bar\F_\sigma=\Pi(\g\otimes\Omega_W(\sigma))\oplus \Pi^{|\sigma|}\g^*$$
lacking canonical odd simplectic structure for $|\sigma|>0$. We use
notation $|\sigma|$ for the dimension of $\sigma$; symbol
$\Pi^{|\sigma|}$ means ``reverse parity if $\sigma$ is
odd-dimensional''. But instead we can think of $\bar{S}_\sigma$ as
of honest effective $BF$ action on simplex $\sigma$, which according
to (\ref{reduction from simplicial complex to simplex}) equals
$S_\sigma(\omega',p';\hbar)=\sum_{\sigma_1\in\sigma}
\bar{S}_{\sigma_1}(\omega'|_{\sigma_1},p'_{(\sigma_1)};\hbar)$,
restricted from full space of infrared fields $\F'_{\sigma}=\Pi
T^*(\Pi(\g\otimes\Omega_W(\sigma)))=\Pi(\g\otimes\Omega_W(\sigma))\oplus[\Omega_W(\sigma)]^*\otimes\g^*$
to the subspace $\bar\F_\sigma\subset\F'_\sigma$, so that
$\bar{S}_\sigma=S_\sigma|_{\bar{F}_\sigma}$. Thus we may name
$\bar{S}_\sigma$ the ``reduced effective action'' on simplex
$\sigma$.

\subsection{Simple cases of elementary BV integral on simplex: dimensions 0 and 1}
\label{calculating S for D=1} Task of computing (\ref{elementary BV
integral on simplex}) on 0-dimensional simplex $\sigma=\Delta^0$ is
trivial, since the space of ultraviolet fields $\Pi
T^*(\Pi(\g\otimes\Omega''(\sigma,\sigma)))$ is empty in this case.
Infrared fields are $\omega'=\omega'_{(0)}=\omega^0\chi_0$,
$p'=p'_{(0)}=e^0 p_0$ and the coordinates $\omega^0\in\Pi\g$,
$p_0\in\g^*$. Hence
$$\bar S(\omega^0,p_0)=<p_0,\frac{1}{2}[\omega^0,\omega^0]>_\g$$
which is indeed a extended $BF$ action on a point. Here
$<\bullet,\bullet>_\g$ is the canonical pairing between $\g$ and
$\g^*$.

Let us now turn to case of dimension 1 for (\ref{elementary BV
integral on simplex}). The 1-dimensional case $\sigma=\Delta^1$
turns out to be exactly solvable, due to the fact that on the
Lagrangian submanifold $\LL_K$ the action we are integrating in
(\ref{elementary BV integral on simplex}) becomes quadratic in
ultraviolet fields.

Whitney forms on interval $\Delta^1$ are: $\chi_0=t_0$,
$\chi_1=t_1$, $\chi_{01}=t_0 dt_1-t_1 dt_0=dt_1$.
 Let us expand infrared
fields as \be\omega'=\omega'_{(0)}+\omega'_{(1)}+\omega'_{(01)}=
\omega^0\chi_0+\omega^1\chi_1+\omega^{01}\chi_{01}\mbox{ and
}p'=p'_{(01)}=e^{01} p_{01}\label{IR fields on interval}\ee with the
coordinates $\omega^0,\omega^1\in\Pi\g$, $\omega^{01}\in\g$ and
$p_{01}\in\Pi\g^*$. Let us also expand ultraviolet fields according
to de Rham degree:
$$\omega''_{(01)}=\omega''^{0}_{(01)}+\omega''^{1}_{(01)}\mbox{ and }
p''_{(01)}=p''^{0}_{(01)}+p''^{1}_{(01)}$$ In these notations the
superscript is the degree of form (or degree of coform for $p$),
while the subscript is the simplex where the ultraviolet field is
supported. Spaces where these components of ultraviolet fields
belong are:
\begin{eqnarray*}\omega''^{0}_{(01)}&\in&\Pi\g\otimes\Omega''^{0}(\Delta^1,\Delta^1)\\
\omega''^{1}_{(01)}&\in&\g\otimes\Omega''^{1}(\Delta^1,\Delta^1)\\
p''^{0}_{(01)}&\in&[\Omega''^{0}(\Delta^1,\Delta^1)]^*\otimes\g^*\\
p''^{1}_{(01)}&\in&[\Omega''^{1}(\Delta^1,\Delta^1)]^*\otimes\Pi\g^*
\end{eqnarray*}
Thus $\omega''^0_{(01)}$ is a $\Pi\g$-valued function on interval
$\Delta^1$ vanishing on the end-points, $\omega''^1_{01}$ is a
$\g$-valued 1-form on $\Delta^1$ with vanishing integral over
$\Delta^1$, $p''^0_{(01)}$ is a $\g^*$-valued 0-coform whose pairing
with linear functions $\chi_0$, $\chi_1$ on $\Delta^1$ vanishes,
$p''^1_{01}$ is a $\Pi\g^*$-valued 1-coform whose pairing with
constant 1-form $\chi_{01}$ vanishes.

Let us choose the homogeneous coordinate $t=t_1$, associated with
right end-point of the interval as the parameter along $\Delta^1$.
The chain homotopy operator (\ref{Dupont's K}) vanishes on functions
$\alpha\in\Omega^0(\Delta^1)$ and acts on 1-forms
$\alpha=\alpha(t)dt\in\Omega^1(\Delta^1)$ as
\begin{multline}K(\alpha(t)dt)=\chi_0 h^0(\alpha)+\chi_1 h^1(\alpha)=\\=t_0 t_1\int_0^1 du\;
\alpha(u t_1)+t_1 (t_1-1)\int_0^1 du\; \alpha(1-u (1-t_1))
=\int_0^{t}dt\;\alpha(t)-t \int_0^1dt\;\alpha(t)\label{K on
interval}\end{multline} It is clearly seen from here that a form on
interval is sent to zero by $K$, iff either it is a 0-form or a
constant 1-form:
$\{\alpha\in\Omega(\Delta^1):\;K\alpha=0\}\;=\Omega^0(\Delta^1)\oplus\Omega^1_W(\Delta^1)$.
Thus the Lagrangian submanifold $\LL_K$ is in our case
\begin{equation}
\LL_K:\left\{\begin{array}{l}\omega''^1_{(01)}=0 \\
p''^0_{(01)}=0\end{array}\right.\label{Lagrangian submanifold D=1}
\end{equation}
Let us expand the action under integral in (\ref{elementary BV
integral on simplex}) on submanifold (\ref{Lagrangian submanifold
D=1}):
\begin{multline} S|_{\LL_K}=<p'_{(01)},d(\omega'_{(0)}+\omega'_{(1)})+
[\omega'_{(0)}+\omega'_{(1)},\omega'_{(01)}]>+<p'_{(01)},[\omega'_{(01)},\omega''^0_{(01)}]>+\\
+<p''^1_{(01)},[\omega'_{(0)}+\omega'_{(1)},\omega'_{(01)}]>+
<p''^1_{(01)},[\omega'_{(01)},\omega''^0_{(01)}]>\label{S on
interval expanded}
\end{multline}
Elementary BV integral (\ref{elementary BV integral on simplex}) can
be interpreted as an integral of type (\ref{BV integral}), inducing
effective action for extended $BF$ theory on one simplex $\sigma$
(i.e. a simplicial complex consisting of $\sigma$ and all its
faces), and then restricting infrared field $p'$ to infrared coforms
of highest degree. Thus we can use perturbative series
(\ref{S'^0},\ref{S'^1}) for (\ref{elementary BV integral on
simplex}). Absence of cubic terms in ultraviolet fields in (\ref{S
on interval expanded}) drastically reduces the number of possible
Feynman diagrams for $\bar S(\omega',p';\hbar)$, and series
(\ref{S'^0},\ref{S'^1}) for tree and 1-loop parts of $\bar S$ are
simplified to
\begin{multline}
\bar S^{(0)}(\omega',p')=<p'_{(01)},d(\omega'_{(0)}+\omega'_{(1)})+
[\omega'_{(0)}+\omega'_{(1)},\omega'_{(01)}]-\\-[K[\omega'_{(0)}+\omega'_{(1)},\omega'_{(01)}],\omega'_{(01)}]
+[K[K[\omega'_{(0)}+\omega'_{(1)},\omega'_{(01)}],\omega'_{(01)}],\omega'_{(01)}]-\\-
[K[K[K[\omega'_{(0)}+\omega'_{(1)},\omega'_{(01)}],\omega'_{(01)}],\omega'_{(01)}],\omega'_{(01)}]+\cdots>
\label{S^0 on interval}
\end{multline}
and
\begin{multline} \bar S^{(1)}(\omega')=-\Str
\;K[\omega'_{(01)},\bullet]+\frac{1}{2}\Str
\;K[\omega'_{(01)},K[\omega'_{(01)},\bullet]]-\\-\frac{1}{3}\Str
\;K[\omega'_{(01)},K[\omega'_{(01)},K[\omega'_{(01)},\bullet]]]+\frac{1}{4}\Str
\;K[\omega'_{(01)},K[\omega'_{(01)},K[\omega'_{(01)},K[\omega'_{(01)},\bullet]]]]-\cdots
\label{S^1 on interval}
\end{multline}
where the super traces are taken on
$\Pi\g\otimes\Omega''^0(\Delta^1,\Delta^1)$ or equivalently on the
whole space of $\Pi\g$-valued 0-forms
$\Pi\g\otimes\Omega^0(\Delta^1)$ (since the diagonal matrix elements
of operators under super-traces vanish on Whitney 0-forms). The
series (\ref{S^0 on interval},\ref{S^1 on interval}) are evaluated
using the two following lemmata.
\begin{Lemma} On 1-dimensional simplex $\Delta^1$, for $n\geq 1$
\be [K(\chi_{01}\wedge\bullet)]^n\circ
\chi_1=-[K(\chi_{01}\wedge\bullet)]^n\circ
\chi_0=\frac{B_{n+1}(t)-B_{n+1}}{(n+1)!}\label{Lemma1 iterated K}\ee
where $B_n(t)$ is the $n$-th Bernoulli polynomial and $B_n=B_n(0)$
is $n$-th Bernoulli number. Also \be
\int_{\Delta^1}\chi_{01}[K(\chi_{01}\wedge\bullet)]^n\circ
\chi_1=-\int_{\Delta^1}\chi_{01}[K(\chi_{01}\wedge\bullet)]^n\circ
\chi_0=-\frac{B_{n+1}}{(n+1)!} \label{Lemma1 integral}\ee
\end{Lemma}
{\bf Proof.} Consider the generating function \be
f(x,t)=\sum_{n=0}^\infty x^n
[K(\chi_{01}\wedge\bullet)]^n\circ\chi_1\label{Lemma 1 gen.
function}\ee Applying $x K(\chi_{01}\wedge\bullet)$ to both sides
and using (\ref{K on interval}) we get the integral equation
$$x \left(\int_0^{t} f dt-t\int_0^1 f dt\right)=f-t$$
and differentiating it w.r.t. $t$ we obtain
$$\frac{\dd}{\dd t} f-1=x\left(f-\int_0^1 fdt\right)$$
and hence $\frac{\dd}{\dd t} f=x f+C(x)$ where $C(x)$ is something
not depending on $t$. Solving this as a differential equation in
variable $t$ with boundary conditions $f(x,0)=0$, $f(x,1)=1$
(emerging from $n=0$ term in (\ref{Lemma 1 gen. function}), the
other terms are vanishing on end-points of interval) yields unique
solution
$$f(x,t)=\frac{e^{x t}-1}{e^x-1}$$
Since Bernoulli polynomials are defined by $$\sum_{n=0}^\infty
\frac{B_n(t)}{n!}x^n=\frac{x e^{xt}}{e^x-1}$$ we obtain
$$[K(\chi_{01}\wedge\bullet)]^n\circ\chi_1=\frac{B_{n+1}(t)-B_{n+1}}{(n+1)!}$$
Fact that $K(\chi_{01}\wedge\bullet)]^n\circ
\chi_1=-[K(\chi_{01}\wedge\bullet)]^n\circ \chi_0$ is obvious from
$K(\chi_{01}\wedge\bullet)]^n\circ
\chi_1+[K(\chi_{01}\wedge\bullet)]^n\circ
\chi_0=[K(\chi_{01}\wedge\bullet)]^n\circ 1=0$. Formula (\ref{Lemma1
integral}) follows directly from (\ref{Lemma1 iterated K}) and from
the following property of Bernoulli polynomials: $\int_0^1 dt\;
B_n(t)=0$ for $n\geq 1$. $\square$

\begin{Lemma} On 1-dimensional simplex $\Delta^1$ for $n\geq 2$
\be \Str_{\Omega^0(\Delta^1)}\;
[K(\chi_{01}\wedge\bullet)]^n=-\frac{B_n}{n!}\label{traces on
interval}\ee
\end{Lemma}

{\bf Proof.} Let us calculate these super-traces (which are now just
ordinary traces, as $\Omega^0(\Delta^1)$ is purely even vector
space) in monomial basis $1,t,t^2,t^3,\ldots\in\Omega^0(\Delta^1)$.
Denote for brevity the operator under super-trace in (\ref{traces on
interval}) by $\MM^n$ with $\MM=K(\chi_{01}\wedge\bullet)$  (letter
$\MM$ for monodromy). For small $n$ we may calculate (\ref{traces on
interval}) directly by finding all diagonal matrix elements on
$\MM^n$. Iterating operator $\MM$ on a monomial $t^m$ we get:
\begin{multline}t^m\stackrel{\MM}{\longrightarrow}\frac{t^{m+1}}{m+1}-\frac{t}{m+1}
\stackrel{\MM}{\longrightarrow}
\\ \stackrel{\MM}{\longrightarrow}\frac{t^{m+2}}{(m+1)(m+2)}-\frac{t^2}{2(m+1)}+
\left(\frac{1}{2(m+1)}-\frac{1}{(m+1)(m+2)}\right)t\stackrel{\MM}{\longrightarrow}\cdots
\label{M on monomials}
\end{multline}
It is clear that for general $m,n$ structure of $\MM^n(t^m)$ is:
$\MM^n(t^m)=\frac{m!}{(m+n)!}t^{m+n}+P_n(t;m)$ where $P_n(t;m)$ is
some polynomial of degree $n$ in $t$ with coefficients being some
rational functions of $m$. Thus all matrix elements
$<t^m|\MM^n|t^m>$ vanish for $m>n$ and only first few contribute to
$\Str$, i.e. those with $1\leq m\leq n$. For instance for $n=2$ from
(\ref{M on monomials}) we obtain
\begin{multline}\Str\;\MM^2=<t|\MM^2|t>+<t^2|\MM^2|t^2>=\\=\left(\frac{1}{2(m+1)}-\frac{1}{(m+1)(m+2)}\right)_{m=1}+
\left(-\frac{1}{2(m+1)}\right)_{m=2}=\left(\frac{1}{2\cdot
2}-\frac{1}{2\cdot 3}\right)+\left(-\frac{1}{2\cdot
3}\right)=-\frac{1}{12}
\end{multline}
Yet for general $n$ we need to calculate somehow the diagonal matrix
elements, and for this we need a generalization of generating
function (\ref{Lemma 1 gen. function}): \be
f_m(x,t)=\sum_{n=0}^\infty x^n [K(\chi_{01}\wedge\bullet)]^n\circ
t^m\label{Lemma 2 gen. function} \ee We again obtain a differential
equation for $f_m$: $$\frac{\dd}{\dd t}f_m=x f_m+m t^{m-1}+C_m(x)$$
where $C_m(x)$ is something not depending on $t$. This equation with
boundary conditions $f_m(x,0)=0$, $f_m(x,1)=1$ uniquely determine
the solution
\begin{multline}
f_m(x,t)=\frac{e^{xt}-1}{e^x-1}\left(1-e^x\int_0^1 d\bar{t}\;
m\bar{t}^{m-1}e^{-x\bar{t}}\right)+e^{xt}\int_0^t
d\bar{t}\;m\bar{t}^{m-1}e^{-x\bar{t}}=\\
=\frac{e^{xt}-1}{e^x-1}\;\sum_{k=0}^{m-1}\frac{m!}{(m-k)!}\;x^{-k}-
\sum_{k=1}^{m-1}\frac{m!}{(m-k)!}\;t^{m-k}x^{-k}\label{Lemma 2
gen.fun. expansion}
\end{multline}
Let us expand $f_m(x,t)$ in powers of $t$:
$f_m(x,t)=\sum_{k=1}^\infty f_{m,k}(x)$. Extracting coefficient of
$t^m$ from $f_m(x,t)$ we obtain the generation function for diagonal
elements of powers of $\MM$ in the following sense:
$$f_{m,m}(x)=<t^m|1|t^m>+x<t^m|\MM|t^m>+x^2<t^m|\MM^2|t^m>+x^3<t^m|\MM^3|t^m>\cdots$$
From the explicit formula (\ref{Lemma 2 gen.fun. expansion}) we have
$$f_{m,m}(x)=1-\frac{1}{e^x-1}\sum_{k=m+1}^\infty\frac{x^k}{k!}$$
The unit term here is the matrix element of identity. Now, to get
generating function for super-traces, we must evaluate the sum
$\sum_{m=1}^\infty (f_{m,m}(x)-1)$:
\begin{multline}
\sum_{m=1}^\infty
(f_{m,m}(x)-1)=x\;\Str\;\MM+x^2\;\Str\;\MM^2+x^3\;\Str\;\MM^3+\cdots=\\=
-\frac{1}{e^x-1}\sum_{m=1}^\infty\sum_{k=m+1}^\infty\frac{x^k}{k!}=
-\frac{1}{e^x-1}\sum_{k=2}^\infty\frac{k-1}{k!}x^k=1-x-\frac{x}{e^x-1}=
-\frac{1}{2}x-\sum_{n=2}^\infty \frac{B_n}{n!}x^n
\end{multline}
Thus we proved that $\Str\;\MM^n=-\frac{B_n}{n!}$ for $n\geq 2$.
$\square$

Now we have all the ingredients to obtain explicit expression for
$\bar{S}$ on a interval $\Delta^1$: we just have to take series
(\ref{S^0 on interval},\ref{S^1 on interval}), plug there the
decompositions of infrared fields (\ref{IR fields on interval}), and
use formulae (\ref{Lemma1 integral},\ref{traces on interval}). We
should also take into account that the first term in (\ref{S^1 on
interval}) vanishes, since it is proportional to the contraction
$f^b_{ab}=0$ of structure constants of gauge algebra. The result is:
\begin{Theorem}
\label{thm: reduced action on 1-simplex} The reduced effective $BF$
action on 1-dimensional simplex $\Delta^1$, as defined by
(\ref{elementary BV integral on simplex}), is
$$\bar{S}(\omega^0,\omega^1,\omega^{01},p_{01};\hbar)=
\bar{S}^{(0)}(\omega^0,\omega^1,\omega^{01},p_{01})+\hbar\;\bar{S}^{(1)}(\omega^{01})$$
and the tree and 1-loop parts of $\bar{S}$ are:
\begin{multline}
\bar{S}^{(0)}(\omega^0,\omega^1,\omega^{01},p_{01})=\\=
<p_{01},-(\omega^1-\omega^0)+\frac{1}{2}[\omega^0+\omega^1,\omega^{01}]-
\sum_{n=2}^\infty
\frac{B_n}{n!}(\ad_{\omega^{01}})^n(\omega^1-\omega^0)>_\g=\\
=<p_{01},\frac{1}{2}[\omega^0+\omega^1,\omega^{01}]-
\left(\frac{\ad_{\omega^{01}}}{2}\coth\frac{\ad_{\omega^{01}}}{2}\right)\;(\omega^1-\omega^0)>_\g
\label{thm: S^0 on interval}
\end{multline}
and
\begin{equation}
\bar{S}^{(1)}(\omega^{01})=\sum_{n=2}^\infty
\frac{1}{n}\;\frac{B_n}{n!}\tr_\g (\ad_{\omega^{01}})^n=
\tr_\g\;\log\left(\frac{\sinh
\frac{\ad_{\omega^{01}}}{2}}{\frac{\ad_{\omega^{01}}}{2}}\right)
\label{thm: S^1 on interval}
\end{equation}
where $\ad_{\omega^{01}}=[\omega^{01},\bullet]$ is the adjoint
action of $\omega^{01}$, $<\bullet,\bullet>_\g$ is canonical pairing
between $\g$ and $\g^*$, $\tr_\g$ is the trace over $\g$.
\end{Theorem}

\subsubsection*{\bf Remark} We can recognize in (\ref{thm: S^0 on
interval}) a special case of Baker-Campbell-Hausdorff series:
$$\bar{S}^{(0)}(\omega^0,\omega^1,\omega^{01},p_{01})=<p_{01},
\left.\frac{\dd}{\dd\e}\right|_{\e=0}\log\left(e^{-\e\omega^1}e^{-\omega^{01}}e^{\e\omega^0}\right)>_\g$$
where $\e\in\Pi\RR$ is an infinitesimal odd variable. The 1-loop
part of reduced effective action on interval (\ref{thm: S^1 on
interval}) has the following interpretation. The measure it defines
on gauge Lie algebra $\g$ is the pullback of Haar measure $\mu_G$ on
gauge group $G$ w.r.t exponential map $\exp:\g\rightarrow G$
$$e^{\bar{S}^{(1)}(\omega^{01})}\delta\omega^{01}=
\Det_g\left(\frac{\sinh
\frac{\ad_{\omega^{01}}}{2}}{\frac{\ad_{\omega^{01}}}{2}}\right)\delta\omega^{01}=\exp^*\mu_G$$
(see e.g. \cite{Kirillov}).

\subsection{Simplicial $BF$ theory on interval}
The simplest example of simplicial $BF$ theory (apart from trivial
0-dimensional case) is the case when the manifold $M$ is an interval
$M=[0,1]$ and triangulation $\Xi$ consists of one 1-dimensional
simplex $[01]$ --- the interval itself and two 0-dimensional
simplices $[0]$, $[1]$ --- the end-points of interval. Infrared
fields are
$\omega'=\omega^0\chi_0+\omega^1\chi_1+\omega^{01}\chi_{01}$ and
$p'=e^0 p_0+e^1 p_1+e^{01} p_{01}$. Here
$\omega^0,\omega^1\in\Pi\g$, $\omega^{01}\in\g$, $p_0,p_1\in\g^*$,
$p_{01}\in\Pi\g^*$. Space of fields $\omega'$ may be identified with
space $\Pi\g\otimes C^*(\Xi)$ of $\Pi\g$-valued cochains on $\Xi$
and space of fields $p'$ with space $C_*(\Xi)\otimes\g^*$ of
$\g^*$-valued chains. The effective action is
\begin{multline}S_{\Xi}(\omega',p';\hbar)=\bar{S}_{0}(\omega^0,p_0)+\bar{S}_{1}(\omega^1,p_1)+
\bar{S}_{01}(\omega^0,\omega^1,\omega^{01},p_{01};\hbar)=\\
=<p_0,\frac{1}{2}[\omega^0,\omega^0]>_g+<p_1,\frac{1}{2}[\omega^1,\omega^1]>_g+\\+
<p_{01},\frac{1}{2}[\omega^0+\omega^1,\omega^{01}]-
\left(\frac{\ad_{\omega^{01}}}{2}\coth\frac{\ad_{\omega^{01}}}{2}\right)\;(\omega^1-\omega^0)>_\g+
\hbar\;\tr_\g\;\log\left(\frac{\sinh
\frac{\ad_{\omega^{01}}}{2}}{\frac{\ad_{\omega^{01}}}{2}}\right)
\end{multline}
Action $S_\Xi$ satisfies quantum master equation by construction and
thus defines cohomological vector field on $\Pi\g\otimes C^*(\Xi)$:
\begin{multline}
Q(\omega')=<\frac{1}{2}[\omega^0,\omega^0],\frac{\dd}{\dd\omega^0}>_\g+
<\frac{1}{2}[\omega^1,\omega^1],\frac{\dd}{\dd\omega^1}>_\g+\\+
<\frac{1}{2}[\omega^0+\omega^1,\omega^{01}]-
\left(\frac{\ad_{\omega^{01}}}{2}\coth\frac{\ad_{\omega^{01}}}{2}\right)\;
(\omega^1-\omega^0),\frac{\dd}{\dd\omega^{01}}>_\g
\end{multline}
Vector field $Q$ generates $L_\infty$ algebra structure on the space
of $\g$-valued cochains $\g\otimes C^*(\Xi)$. The 1-loop part of
action $S_\Xi$ produces density function $\rho(\omega')$ on
$\Pi\g\otimes C^*(\Xi)$:
$$\rho(\omega')=e^{S_\Xi^{(1)}(\omega')}=\Det_g \left(\frac{\sinh
\frac{\ad_{\omega^{01}}}{2}}{\frac{\ad_{\omega^{01}}}{2}}\right)$$
Density $\rho$ is $Q$-invariant by construction. The $L_\infty$
quasi-isomorphism $U: \Pi\g\otimes C^*(\Xi)\rightarrow
\Pi\g\otimes\Omega(\Delta^1)$ is easily found from (\ref{series for
U}) using (\ref{Lemma1 iterated K}):
$$
U(\omega')=\omega^0+\left(\frac{1-e^{-t\;\ad_{\omega^{01}}}}{1-e^{-\ad_{\omega^{01}}}}
\right)(\omega^1-\omega^0)+\omega^{01}dt$$ or  equivalently in more
symmetric form:
$$U(\omega')=\left(\frac{1-e^{t_0\;\ad_{\omega^{01}}}}{1-e^{\ad_{\omega^{01}}}}
\right)\omega^0+\left(\frac{1-e^{-t_1\;\ad_{\omega^{01}}}}{1-e^{-\ad_{\omega^{01}}}}
\right)\omega^1+\omega^{01}dt$$

Let us now take a triangulation $\Xi$ on the interval $I=[0,1]$,
consisting of $N\geq 1$ 1-dimensional simplices
$[01],[12],\ldots,[(N-1)N]$ and $N+1$ 0-dimensional simplices
$[0],[1],\ldots,[N]$. Let the coordinate of $[i]$ on the interval
$[0,1]$ be $\frac{i}{N}$, and let $\e=\frac{1}{N}$ denote the
spacing. The infrared fields are: $\omega'=\sum_{i=0}^N \omega^i
\chi_i+\sum_{i=1}^N \omega^{i-1,i}\chi_{i-1,i}$, $p'=\sum_{i=0}^N
e^i p_i+\sum_{i=1}^N e^{i-1,i} p_{i-1,i}$ and the effective action
is $$S_\Xi(\omega',p';\hbar)=\sum_{i=0}^N \bar{S}_{i}(\omega^i,p_i)+
\sum_{i=1}^N\bar{S}_{i-1,i}(\omega^{i-1},\omega^i,\omega^{i-1,i},p_{i-1,i};\hbar)$$
Let us introduce normalized coordinates on space of infrared fields:
$\omega^i=\tilde\omega^i$,
$\omega^{i-1,i}=\e\,\tilde\omega^{i-1,i}$, $p_i=\e\,\tilde p_i$,
$p_{i-1,i}=\tilde{p}_{i-1,i}$. Then the projector $\PP'$ acts on
smooth forms $\omega(t)\in\Pi\g\otimes\Omega([0,1])$ as
$$\tilde\omega^i=\omega(\frac{i}{N}),\;\tilde\omega^{i-1,i}=
\frac{1}{\e}\int_{\frac{i-1}{N}}^{\frac{i}{N}}\omega$$ Thus
$\tilde\omega$ is what we would call a ``lattice approximation'' of
a smooth form. Expressing effective action $S_\Xi$ in terms of these
normalized infrared fields, we obtain:
\begin{multline}
S_\Xi(\omega',p';\hbar)=\\\e\left(-\sum_{i=1}^N
<\tilde{p}_{i-1,i},\frac{\tilde\omega^i-\tilde\omega^{i-1}}{\e}>_\g+
\sum_{i=0}^N
<\tilde{p}_i,\frac{1}{2}[\tilde\omega^i,\tilde\omega^i]>_\g+
\sum_{i=1}^N<\tilde{p}_{i-1,i},[\frac{\tilde\omega^{i-1}+\tilde\omega^{i}}{2},\tilde\omega^{i-1,i}]>_\g\right)-\\-
\sum_{n=2}^\infty \e^{n+1}\frac{B_{n}}{n!}\sum_{i=1}^N
<p_{i-1,i},(\ad_{\tilde\omega^{i-1,i}})^n
\left(\frac{\tilde\omega^i-\tilde\omega^{i-1}}{\e}\right)>_\g
+\hbar\sum_{n=2}^\infty\e^n \frac{B_n}{n\;n!}\sum_{i=1}^N\tr_\g
(\ad_{\tilde\omega^{i-1,i}})^n \label{lattice action on interval}
\end{multline}
Expression (\ref{lattice action on interval}) constitutes a subtle
lattice version of ordinary extended $BF$ action (on smooth forms)
on interval $S=<p,d\omega+\frac{1}{2}[\omega,\omega]>$. The first
three terms in (\ref{lattice action on interval}) are what we would
call the naive lattice action for extended $BF$ theory on interval,
while the other terms are corrections of higher order in spacing
$\e$. These additional terms make this lattice action satisfy
quantum master equation (the naive lattice action does not satisfy
QME).

\subsection{Simplicial $BF$ theory on circle, induction to cohomologies of circle,
$BF$ state-sum on circle} Now let $M=S^1$ be a circle, and $\Xi$ be
a triangulation of circle, consisting of $N\geq 2$ 1-simplices
$[01], [12],\ldots,[(N-1)N]$ and $N$ 0-simplices
$[1],[2],\dots,[N]$. Infrared fields are
$\omega'=\sum_{i=1}^N\omega^i\chi_i+\sum_{i=1}^N\omega^{i-1,i}\chi_{i-1,i}$,
$p'=\sum_{i=1}^N e^i p_i+\sum_{i=1}^N e^{i-1,i} p_{i-1,i}$ and
effective action is
\begin{multline}
S_\Xi(\omega',p';\hbar)=\sum_{i=1}^N<p_i,\frac{1}{2}[\omega^i,\omega^i]>_\g+\\+
\sum_{i=1}^N<p_{i-1,i},\frac{1}{2}[\omega^{i-1}+\omega^i,\omega^{i-1,i}]-
\left(\frac{\ad_{\omega^{i-1,i}}}{2}\coth\frac{\ad_{\omega^{i-1,i}}}{2}\right)\;(\omega^i-\omega^{i-1})>_\g+\\+
\hbar\sum_{i=1}^N\tr_\g\;\log\left(\frac{\sinh
\frac{\ad_{\omega^{i-1,i}}}{2}}{\frac{\ad_{\omega^{i-1,i}}}{2}}\right)\label{S
on circle}
\end{multline}
We use here the convention $\omega^0=\omega^{N}$.

Let us induce effective action on cohomologies of circle from action
(\ref{S on circle}), considered as $BF_\infty$ type theory. For
simplicity take $N=2$, i.e. the simplest non-degenerate
triangulation of the circle. Next we split the fields living on
triangulation $\Xi$ into (new) infrared and ultraviolet parts:
\begin{eqnarray*}
\omega^1=\omega^A-\frac{1}{2}\omega''^A,\;
\omega^2=\omega^A+\frac{1}{2}\omega''^A,\;
\omega^{01}=\frac{\omega^B-\omega''^B}{2},\;
\omega^{12}=\frac{\omega^B+\omega''^B}{2},\;\\
p_1=\frac{p_A-p''_A}{2},\; p_2=\frac{p_A+p''_A}{2},\;
p_{01}=p_B-\frac{1}{2}p''_B,\; p_{12}=p_B+\frac{1}{2}p''_B
\end{eqnarray*}
Here $A$ and $B$ label the 0- and 1-dimensional cohomologies of
circle: $e_A=1$, $e_B=dt$. The Lagrangian submanifold $\LL_K$ is:
$\omega''^B=0$, $p''_A=0$. Restricted to $\LL_K$ the action (\ref{S
on circle}) gives
\begin{multline}S_\Xi|_{\LL_K}=<p_A,\frac{1}{2}[\omega^A,\omega^A]+
\frac{1}{8}[\omega''^A,\omega''^A]>_\g+<p_B,[\omega^A,\omega^B]>_\g-\\-
<p''_B,\left(\frac{\ad_{\omega^B}}{4}\coth\frac{\ad_{\omega^B}}{4}\right)\omega''^A>_\g+
\hbar\;\tr_g\log\left(\frac{\sinh
\frac{\ad_{\omega^B}}{4}}{\frac{\ad_{\omega^B}}{4}}\right)^2
\end{multline}
The BV integral (\ref{BV integral}) is
\begin{multline}e^{\frac{1}{\hbar}S_{H^*(S^1)}}=\int e^{S_\Xi/\hbar}\;\delta\omega''^A\delta
p''_B=\exp\left(\frac{1}{\hbar}(<p_A,\frac{1}{2}[\omega^A,\omega^A]>_\g+<p_B,[\omega^A,\omega^B]>_\g)\right)\cdot\\
\cdot\Det_\g\left(\frac{\ad_{\omega^B}}{4}\coth\frac{\ad_{\omega^B}}{4}\right)\;
\Det_g\left(\frac{\sinh
\frac{\ad_{\omega^B}}{4}}{\frac{\ad_{\omega^B}}{4}}\right)^2=\\
=\exp\left(\frac{1}{\hbar}(<p_A,\frac{1}{2}[\omega^A,\omega^A]>_\g+<p_B,[\omega^A,\omega^B]>_\g)\right)\;
\Det_g\left(\frac{\sinh
\frac{\ad_{\omega^B}}{2}}{\frac{\ad_{\omega^B}}{2}}\right)
\end{multline}
Hence the effective action on cohomologies of circle (i.e. on space
$\Pi T^* (\Pi\g\otimes H^*(S^1))$) is
\begin{multline}
S_{H^*(S^1)}(\omega^A,\omega^B,p_A,p_B;\hbar)=\\=
<p_A,\frac{1}{2}[\omega^A,\omega^A]>_\g+<p_B,[\omega^A,\omega^B]>_\g+
\hbar\;\tr_g\log\left(\frac{\sinh
\frac{\ad_{\omega^B}}{2}}{\frac{\ad_{\omega^B}}{2}}\right)\label{S
on cohomologies of circle}
\end{multline}
This action coincides with (\ref{S on circle}) if we formally set
$N=1$ (although this corresponds to degenerate triangulation).
Indeed we could derive (\ref{S on cohomologies of circle}) directly
from continuous extended $BF$ theory on circle and arrive to the
same answer. The difference is that inducing (\ref{S on cohomologies
of circle}) from (\ref{S on circle}) we only need to calculate a
finite-dimensional integral. Action (\ref{S on cohomologies of
circle}) generates Lie algebra structure on $\g\otimes H^*(S^1)$ and
no higher homotopic operations (Massey operations), since circle is
a formal manifold. The 1-loop part of (\ref{S on cohomologies of
circle}) gives $Q$-invariant density function on $\Pi\g\otimes
H^*(S^1)$:
$$\rho(\omega^B)=\Det_g\left(\frac{\sinh
\frac{\ad_{\omega^B}}{2}}{\frac{\ad_{\omega^B}}{2}}\right)$$

The state-sum for extended $BF$ theory on circle, according to
definition from (\ref{state-sum def}) is then \be
Z(\g\otimes\Omega(S^1))=\int_{\Pi
H^*(S^1)}\rho(\omega^B)\;\delta\omega^A\delta\omega^B=
\int_\g\Det_g\left(\frac{\sinh
\frac{\ad_{\omega^B}}{2}}{\frac{\ad_{\omega^B}}{2}}\right)\delta\omega^B\label{Z
on circle}\ee As we observed before, the measure we are integrating
is a pullback of Haar measure on the gauge Lie group $G$ under
exponential map $\exp:\g\rightarrow G$. Notice that integral (\ref{Z
on circle}) (if taken over the connected component of support of
$\rho$, containing zero
--- the integral over whole $\g$ diverges) gives the volume of the
gauge group:
$$Z(\g\otimes\Omega(S^1))=\vol(G)$$
As we mentioned in section \ref{state-sum def}, we do not have a
good explanation, why we should regularize the integral for
state-sum in such a way. Notice also that the compactness of gauge
group suddenly becomes important for finiteness of state-sum. This
should be viewed as an essentially quantum phenomenon.

\subsection{Elementary BV integral on simplex of dimension $D\geq 2$: perturbative results}
Integral (\ref{elementary BV integral on simplex}) on
$D$-dimensional simplex $\Delta^D$ is no longer Gaussian if $D\geq
2$ and we do not know the closed expression for
$\bar{S}_{\Delta^D}$. But we can use perturbative expansion
(\ref{S'^0},\ref{S'^1}) for $\bar{S}_{\Delta^D}$ and calculate its
first terms explicitly.

We use the same notation for Taylor expansion of $\bar{S}$ as in
subsection \ref{effective action for BF_infty}:
$$\bar{S}_{\Delta^D}(\omega,p_{\Delta^D})=
\sum_{n=1}^\infty
<p_{\Delta^D},\frac{1}{n!}\;\bar{l}^{(n)}_{\Delta^D}(\omega,\ldots,\omega)>_\g+
\hbar \sum_{n=1}^\infty
\frac{1}{n!}\;\bar{q}^{(n)}_{\Delta^D}(\omega,\ldots,\omega)$$ where
$\bar{l}^{(n)}_{\Delta^D}$ is a $n$-linear super-antisymmetric map
$$\bar{l}^{(n)}_{\Delta^D}:(\Pi\g\otimes\Omega_W(\Delta^D))^{\otimes
n}\rightarrow [\Omega^D_W(\Delta^D)]^*\otimes\g^*\simeq\Pi^D\g^*$$
and $\bar{q}^{(n)}_{\Delta^D}$ is a $n$-linear super-antisymmetric
function
$$\bar{q}^{(n)}_{\Delta^D}:(\Pi\g\otimes\Omega_W(\Delta^D))^{\otimes
n}\rightarrow\RR$$ and we put bars on $l$ and $q$ to indicate that
they correspond to the reduced effective action on simplex.

Now introduce a set of functions $C_{T}$ on faces of $\Delta^D$ as
follows: for every rooted binary tree $T$ with $|T|=n$ leaves and
$\sigma_1,\ldots,\sigma_n$ faces of $\Delta^D$ we define
$$C_T(\sigma_1,\ldots,\sigma_n)=\int_{\Delta^D}\Iter_{T,\;K(\bt\wedge\bt),\;\bt\wedge\bt}\;
(\chi_{\sigma_1},\ldots,\chi_{\sigma_n})$$ For the trivial tree with
one leaf we set
$$C_{(*)}(\sigma_1)=\int_{\Delta^D}d\chi_{\sigma_1}$$
We also introduce the sign functions $\e_T$ taking values in
$\{-1,0,+1\}$, defined as
$$\e_T(\sigma_1,\ldots,\sigma_n)=\left\{
\begin{array}{ll}
+1&\mbox{if }\;C_T(\sigma_1,\ldots,\sigma_n)>0 \\
0&\mbox{if }\;C_T(\sigma_1,\ldots,\sigma_n)=0 \\
-1&\mbox{if }\;C_T(\sigma_1,\ldots,\sigma_n)<0
\end{array}\right.$$

Functions $C_T$ have the following symmetry properties:
\begin{itemize}
\item Internal symmetry: for $\pi_1,\ldots,\pi_n$ permutations of
vertices of simplices $\sigma_1,\ldots,\sigma_n$ \be
C_T(\pi_1\sigma_1,\ldots\pi_n\sigma_n)=(-1)^{\pi_1}\cdots(-1)^{\pi_n}
C_T(\sigma_1,\ldots,\sigma_n)\label{C_T internal}\ee where
$(-1)^{\pi_i}$ is the sign of permutation $\pi_i$.
\item External symmetry: for $\pi$ a permutation of vertices of $\Delta^D$
\be C_T(\pi\sigma_1,\ldots,\pi\sigma_n)=(-1)^\pi
C_T(\sigma_1,\ldots,\sigma_n)\label{C_T external}\ee
\item Symmetry under tree isomorphisms: if trees $T$ and $T'$ are
isomorphic as non-planar graphs and $\kappa: T\rightarrow T'$ is the
isomorphism, then \be
C_{T'}(\sigma_{\kappa(1)},\ldots,\sigma_{\kappa(n)})=\e_{\kappa}(|\sigma_1|,\ldots,|\sigma_n|)\;
C_T(\sigma_1,\ldots,\sigma_n)\label{C_T under tree isomorphism}\ee
where we understand that $\kappa$ maps leaves of $T$ into leaves of
$T'$. The sign $\e_{\kappa}(|\sigma_1|,\ldots,|\sigma_n|)=\pm 1$
depends only on dimensions of faces, not on faces themselves and is
defined by (\ref{C_T under tree isomorphism}).  Important case of
this symmetry is when $T'=T$ and $\kappa\in\Aut(T)$.
\end{itemize}

Examples of symmetry (\ref{C_T under tree isomorphism}):
\begin{eqnarray*}
C_{(*(**))}(\sigma_3,\sigma_1,\sigma_2)=(-1)^{(|\sigma_1|+|\sigma_2|-1)\;|\sigma_3|}
C_{((**)*)}(\sigma_1,\sigma_2,\sigma_3)\\C_{((**)*)}(\sigma_2,\sigma_1,\sigma_3)=(-1)^{|\sigma_1|\;|\sigma_2|}
C_{((**)*)}(\sigma_1,\sigma_2,\sigma_3)
\end{eqnarray*}
Obviously symmetries (\ref{C_T internal},\ref{C_T external},\ref{C_T
under tree isomorphism}) also hold for sign functions $\e_T$.

\begin{Lemma}\label{Lemma 3} Values of $C_T$ for $|T|\leq 3$ are
given by
\begin{eqnarray}
C_{(*)}(\sigma_1)&=&\e_{(*)}(\sigma_1),\\
C_{(**)}(\sigma_1,\sigma_2)&=&\e_{(**)}(\sigma_1,\sigma_2)\;
\frac{|\sigma_1|!\;|\sigma_2|!}{(|\sigma_1|+|\sigma_2|+1)!},\\
C_{((**)*)}(\sigma_1,\sigma_2,\sigma_3)&=&\e_{((**)*)}(\sigma_1,\sigma_2,\sigma_3)\;
\frac{|\sigma_1|!\;|\sigma_2|!\;|\sigma_3|!}{(|\sigma_1|+|\sigma_2|+1)\;(|\sigma_1|+|\sigma_2|+|\sigma_3|+1)!}
\end{eqnarray}
and signs $\e_{(*)}$, $\e_{(**)}$, $\e_{((**)*)}$ are uniquely
determined by symmetries (\ref{C_T internal},\ref{C_T
external},\ref{C_T under tree isomorphism}), specific values
\begin{eqnarray}
\e_{(*)}([12\cdots D])=1,\label{e_1 standard}\\
\e_{(**)}([0\cdots a],[a\cdots D])=1\mbox{ for }0\leq a\leq D,\label{e_2 standard}\\
\e_{((**)*)}([0\cdots a],[a\cdots a+b],[a(a+b)\cdots
D])=(-1)^{a+b+1}\label{e_3 standard}
\\
\mbox{ for }0\leq a \leq D-1,\; 1\leq b\leq D-a\nonumber
\end{eqnarray}
and non-vanishing conditions
\begin{eqnarray*}
\e_{(*)}(\sigma_1)\neq 0\;\mbox{ iff }|\sigma_1|=D-1,\\
\e_{(**)}(\sigma_1,\sigma_2)\neq 0\;\mbox{ iff
}|\sigma_1|+|\sigma_2|=D\mbox{ and }\sigma_1\cup\sigma_2=\Delta^D,\\
\e_{((**)*)}(\sigma_1,\sigma_2,\sigma_3)\neq 0\mbox{ iff
}|\sigma_1|+|\sigma_2|+|\sigma_3|=D+1,\;\sigma_1\cup\sigma_2\cup\sigma_3=\Delta^D\\
\mbox{ and }
\sigma_1\cap\sigma_2=\sigma_1\cap\sigma_2\cap\sigma_3\mbox{ is a
0-simplex}
\end{eqnarray*}
\end{Lemma}
Here $\cup$ means union of simplices viewed as sets of vertices (or
equivalently convex hull of geometric simplices), $\cap$ means
intersection.

The absolute value of $C_T$ turns out to be a simple function of
dimensions of simplices, non-vanishing condition is a combinatorial
condition formulated in terms of dimensions and unions/intersections
of simplices, while the sign $\e_T$ of $C_T$ is the most tricky
thing here, determined by reduction to standard cases (\ref{e_1
standard},\ref{e_2 standard},\ref{e_3 standard}) via symmetries
(\ref{C_T internal},\ref{C_T external},\ref{C_T under tree
isomorphism}).

We need coefficient functions $C_T$ for evaluating terms of
perturbative expansion (\ref{S'^0}) for the reduced effective
action:
\begin{multline}
<p,(-1)^{|T|}\Iter_{T,\;K[\bt,\bt],\;[\bt,\bt]}(\omega,\ldots,\omega>=\\
=\sum_{\sigma_1,\ldots,\sigma_n\subset\Delta^D} <e^{\Delta^D}
p_{\Delta^D},(-1)^{|T|}\;\Iter_{T,\;K[\bt,\bt],\;[\bt,\bt]}
(\omega^{\sigma_1}\chi_{\sigma_1},\ldots,\omega^{\sigma_n}\chi_{\sigma_n})>=\\=
\sum_{\sigma_1,\ldots,\sigma_n\subset\Delta^D}\tilde\e_T(|\sigma_1|,\ldots,|\sigma_n|)\;
C_T(\sigma_1,\ldots,\sigma_n)<p_{\Delta^D},\Iter_{T,\;[\bt,\bt]}(\omega^{\sigma_1},\ldots,\omega^{\sigma_n})>_\g
\label{S_T via C_T}
\end{multline}
Sign $\tilde\e_T$ comes from interchanging coordinates $\omega$ and
Whitney forms $\chi$ in (\ref{S_T via C_T}), and is defined by
\begin{multline}(-1)^{|T|}\;\Iter_{T,\;K[\bt,\bt],\;[\bt,\bt]}
(\omega^{\sigma_1}\chi_{\sigma_1},\ldots,\omega^{\sigma_n}\chi_{\sigma_n})=\\
=\tilde\e_T(|\sigma_1|,\ldots,|\sigma_n|) \;
C_T(\sigma_1,\ldots,\sigma_n)\;\Iter_{T,\;[\bt,\bt]}(\omega^{\sigma_1},\ldots,\omega^{\sigma_n})
\end{multline}
For trees with $|T|\leq 3$, $\tilde\e_T$ is given by
\begin{eqnarray}
\tilde\e_{(*)}(|\sigma_1|)&=&(-1)^{|\sigma_1|+1},\\
\tilde\e_{(**)}(|\sigma_1|,|\sigma_2|)&=&(-1)^{|\sigma_1|\;(|\sigma_2|+1)},\\
\tilde\e_{((**)*)}(|\sigma_1|,|\sigma_2|,|\sigma_3|)&=&(-1)^{|\sigma_1|\;|\sigma_2|+|\sigma_1|\;|\sigma_3|+
|\sigma_2|\;|\sigma_3|+|\sigma_1|+|\sigma_3|}
\end{eqnarray}
Notice that (\ref{S_T via C_T}) does not depend on planar structure
of tree $T$: if $T$ and $T'$ are isomorphic as non-planar graphs,
then $\bar{S}_{T'}=\bar{S}_T$. At the same time $\tilde\e_T$, $C_T$
and
$\Iter_{T,\;[\bt,\bt]}(\omega^{\sigma_1},\ldots,\omega^{\sigma_n})$
separately do depend on planar structure of $T$.

The expansion (\ref{S'^0}) for tree part of reduced effective action
on simplex $\Delta^D$ in terms of polylinear maps
$\bar{l}^{(n)}_{\Delta^D}$ becomes
\begin{multline}\bar{l}^{(n)}_{\Delta^D}(\omega,\ldots,\omega)=\\=
n!\sum_{T:\;|T|=n}\frac{1}{\Aut(T)}\sum_{\sigma_1,\ldots,\sigma_n\subset\Delta^D}
\tilde\e_T(|\sigma_1|,\ldots,|\sigma_n|)\;
C_T(\sigma_1,\ldots,\sigma_n)\;\Iter_{T,\;[\bt,\bt]}(\omega^{\sigma_1},\ldots,\omega^{\sigma_n})
\end{multline}
where we sum over classes of isomorphic trees (or equivalently over
trees without specified embedding into plane). Using Lemma
\ref{Lemma 3} we obtain explicit expressions for
$\bar{l}^{(n)}_{\Delta^D}$ with $n=1,2,3$ (and thus expansion for
$\bar{S}^{(0)}_{\Delta^D}$ up to order $\OO(p\omega^3)$).
\begin{Theorem}
\label{thm: first classical higher operations} The first terms in
tree part of reduced effective action $\bar{S}^{(0)}_{\Delta^D}$ are
given by
\begin{eqnarray}
\bar{l}^{(1)}_{\Delta^D}(\omega)&=&\sum_{\sigma_1\subset\Delta^D}(-1)^{|\sigma_1|+1}\e_{(*)}(\sigma_1)\;\omega^{\sigma_1},\\
\bar{l}^{(2)}_{\Delta^D}(\omega,\omega)&=&
\sum_{\sigma_1,\sigma_2\subset\Delta^D}(-1)^{|\sigma_1|\;(|\sigma_2|+1)}\e_{(**)}(\sigma_1,\sigma_2)
\;\frac{|\sigma_1|!\;|\sigma_2|!}{(|\sigma_1|+|\sigma_2|+1)!}\;[\omega^{\sigma_1},\omega^{\sigma_2}],\\
\label{l^3}\bar{l}^{(3)}_{\Delta^D}(\omega,\omega,\omega)&=&
3\sum_{\sigma_1,\sigma_2,\sigma_3\subset\Delta^D}(-1)^{|\sigma_1|\;|\sigma_2|+|\sigma_1|\;|\sigma_3|+
|\sigma_2|\;|\sigma_3|+|\sigma_1|+|\sigma_3|}\e_{((**)*)}(\sigma_1,\sigma_2,\sigma_3)
\cdot\\ &&\cdot
\frac{|\sigma_1|!\;|\sigma_2|!\;|\sigma_3|!}{(|\sigma_1|+|\sigma_2|+1)\;(|\sigma_1|+|\sigma_2|+|\sigma_3|+1)!}
\;[[\omega^{\sigma_1},\omega^{\sigma_2}],\omega^{\sigma_3}]\nonumber
\end{eqnarray}
\end{Theorem}

Let us now turn to the 1-loop part of reduced effective action
$\bar{S}^{(1)}_{\Delta^D}(\omega)$. Similarly to what we did for
tree part, for every loop graph $L$ with $|L|=n$ leaves we introduce
a function $C_L$ on faces of $\Delta^D$:
$$C_L(\sigma_1,\ldots,\sigma_n)=\Loop_{L,\;K(\bt\wedge\bt),\;\Omega(\Delta^D)}
(\chi_{\sigma_1},\ldots,\chi_{\sigma_n})$$ where the super-trace is
taken over the space $\Omega(\Delta^D)$ of all differential forms on
$\Delta^D$, and binary operator $K(\bt\wedge\bt)$ is acting on the
same space. Obviously $C_L$ like $C_T$ possesses internal symmetry
(\ref{C_T internal}) and symmetry under graph isomorphisms (\ref{C_T
under tree isomorphism}) and the following form of external
symmetry: for $\pi$ a permutation of vertices of $\Delta^D$ \be
C_L(\pi\sigma_1,\ldots,\pi\sigma_n)=C_L(\sigma_1,\ldots,\sigma_n)\label{external
symmetry for C_L}\ee only difference from the case of trees is the
absence of sign $(-1)^\pi$.

Using $C_L$ we may evaluate the terms of expansion (\ref{S'^1}) for
reduced effective action on $\Delta^D$ as follows
\begin{multline}
(-1)^{|L|}\;\Loop_{L,\;K[\bt,\bt],\;\Pi\g\otimes\Omega(\Delta^D)}(\omega,\ldots,\omega)=\\=
\sum_{\sigma_1,\ldots,\sigma_n\subset\Delta^D}
(-1)^{|L|}\;\Loop_{L,\;K[\bt,\bt],\;\Pi\g\otimes\Omega(\Delta^D)}(\omega^{\sigma_1}\chi_{\sigma_1},
\ldots,\omega^{\sigma_n}\chi_{\sigma_n})=\\=
\sum_{\sigma_1,\ldots,\sigma_n\subset\Delta^D}\tilde\e_L(|\sigma_1|,\ldots,|\sigma_n|)\;
C_L(\sigma_1,\ldots,\sigma_n)\;\Loop_{L,[\bt,\bt],\g}(\omega^{\sigma_1},\ldots,\omega^{\sigma_n})
\label{S_L via C_L}
\end{multline}
The meaning of (\ref{S_L via C_L}) is to separate super-trace over
$\Pi\g\otimes\Omega(\Delta^D)$ into trivial part --- trace over $\g$
and non-trivial part --- super-trace over infinite-dimensional space
$\Omega(\Delta^D)$. Signs $\tilde\e_L$ come from interchanging
$\omega^\sigma$ and $\chi_\sigma$ and are defined by (\ref{S_L via
C_L}). Plugging (\ref{S_L via C_L}) into (\ref{S'^1}), we obtain
\begin{multline}
\bar{q}^{(n)}_{\Delta^D}(\omega,\ldots,\omega)=\\=n!\sum_{L:\;|L|=n}\frac{1}{\Aut(L)}
\sum_{\sigma_1,\ldots,\sigma_n\subset\Delta^D}
\tilde\e_L(|\sigma_1|,\ldots,|\sigma_n|)\;
C_L(\sigma_1,\ldots,\sigma_n)\;\Loop_{L,[\bt,\bt],\g}(\omega^{\sigma_1},\ldots,\omega^{\sigma_n})
\label{q^n via C_L}
\end{multline}
Here we sum over classes of isomorphic 1-loop graphs. Graphs $L$
with cycle of length 1 do not contribute to (\ref{q^n via C_L})
since for these graphs $\Loop_{L,[\bt,\bt],\g}$ is proportional to
the contraction $f^{b}_{ab}$ of structure constants of gauge
algebra, and thus these terms vanish. For instance this means that
$\bar{q}^{(1)}_{\Delta^D}=0$. For $\bar{q}^{(2)}_{\Delta^D}$ the
only contributing graph is $L=(*(*\bullet))$. Symmetries (\ref{C_T
internal},\ref{external symmetry for C_L}) for $C_{(*(*\bullet))}$
allow only two possible terms for $\bar{q}^{(2)}_{\Delta^D}$:
$$\bar{q}^{(2)}_{\Delta^D}(\omega,\omega)=\A_D\sum_{0\leq i<j\leq
D}\tr_\g\;(\ad_{\omega^{ij}})^2+ \B_D\sum_{0\leq i<j<k\leq
D}\tr_\g\; (\ad_{\omega^{jk}}-\ad_{\omega^{ik}}+\ad_{\omega^{ij}})^2
$$
Here $\A_D$ and $\B_D$ are some coefficients, and symmetries tell
nothing of their values. It turns out that value of $\A_D$ can be
recovered from master equation for full effective action on simplex
$\Delta^D$ (i.e. sum of reduced effective actions on all faces) and
result (\ref{l^3}) for tree part of effective action. Coefficient
$\B_D$ on the other hand cannot be recovered from master equation
since the canonical transformation
$$S_{\Delta^D}\mapsto S_{\Delta^D}+\hbar\;\alpha\; Q \left(\sum_{0\leq i<j<k\leq D}\tr_\g\;
(\ad_{\omega^{jk}}-\ad_{\omega^{ik}}+\ad_{\omega^{ij}})\cdot\ad_{\omega^{ijk}}\right)$$
shifts coefficient $\B_D$ by $\alpha$ (and gives indeed a solution
to master equation). This also means that coefficient $\B_D$ is
somehow less important then $\A_D$, since it stands in front of
$Q$-exact term.

\begin{Theorem}
\label{thm: first quantum operation} The first terms of 1-loop part
of reduced effective action $\bar{S}^{(1)}_{\Delta^D}$ are given by
$$
\bar{q}^{(1)}_{\Delta^D}(\omega)=0$$ and \be
\bar{q}^{(2)}_{\Delta^D}(\omega,\omega)=\A_D\sum_{0\leq i<j\leq
D}\tr_\g\;(\ad_{\omega^{ij}})^2+ \B_D\sum_{0\leq i<j<k\leq
D}\tr_\g\; (\ad_{\omega^{jk}}-\ad_{\omega^{ik}}+\ad_{\omega^{ij}})^2
\label{q^2 allowed terms}\ee and coefficient $\A_D$ is \be
\A_D=\frac{(-1)^{D+1}}{(D+1)^2\;(D+2)}\label{q^2 first
coefficient}\ee
\end{Theorem}

We also carried out an explicit calculation of super-trace
$C_{(*(*\bt))}$ in dimensions $D=2,3$ (not relying on master
equation arguments) and found out the following:
\begin{Theorem}
\label{operation 2->0 in D=2, D=3} In dimensions $D=2,3$ the
lowest-order term in 1-loop part of reduced effective action
$\bar{q}^{(2)}_{\Delta^D}$ is given by (\ref{q^2 allowed terms})
with
\begin{eqnarray*}
\A_2=-\frac{1}{36},\;\B_2=\frac{1}{270},\\
\A_3=\frac{1}{80},\;\B_3=-\frac{1}{648}
\end{eqnarray*}
\end{Theorem}
For the coefficient $\B_D$ we might try to guess some formula like
$$\B_D=\frac{(-1)^D}{9\,D(D+1)(D+3)}$$
from here, but this is just a guess, and our evidence is limited to
only two points $D=2,3$.

Collecting our results for $D=2$ we obtain
\begin{multline}
\bar{S}_{[012]}(\omega,p;\hbar)=
\\
=<p_{012},(\omega^{01}+\omega^{12}+\omega^{20})+
\frac{1}{3}[\omega^0+\omega^1+\omega^2,\omega^{012}]+\frac{1}{6}([\omega^{01},\omega^{12}]+[\omega^{12},\omega^{20}]+
[\omega^{20},\omega^{12}])+\\
+\frac{1}{72}([[\omega^{01}+\omega^{12}+\omega^{20},\omega^{01}],\omega^{01}]+
[[\omega^{01}+\omega^{12}+\omega^{20},\omega^{12}],\omega^{12}]+
[[\omega^{01}+\omega^{12}+\omega^{20},\omega^{20}],\omega^{20}])-
\\
-\frac{1}{24}([[\omega^1-\omega^0,\omega^{01}],\omega^{012}]+
[[\omega^2-\omega^1,\omega^{12}],\omega^{012}]+
[[\omega^0-\omega^2,\omega^{20}],\omega^{012}])-\\-
\frac{1}{36}([[\omega^1-\omega^0,\omega^{012}],\omega^{01}]+
[[\omega^2-\omega^1,\omega^{012}],\omega^{12}]+
[[\omega^0-\omega^2,\omega^{012}],\omega^{20}])>_\g+\\
+\hbar\;\tr_g
\left(-\frac{1}{72}\;((\ad_{\omega^{01}})^2+(\ad_{\omega^{12}})^2+(\ad_{\omega^{20}})^2)
+\frac{1}{540}(\ad_{\omega^{01}}+\ad_{\omega^{12}}+\ad_{\omega^{20}})^2\right)+\\+
\OO(p\,\omega^4)+\OO(\hbar\,\omega^3)
\end{multline}
Importance of effective action on 2-simplex is that its tree part
restricted to Whitney 1-forms produces a formula for ``simplicial
curvature'' of simplicial (i.e. Whitney) connection 1-form:
\begin{multline}
F_{[012]}(\omega^{01},\omega^{12},\omega^{20})=
(\omega^{01}+\omega^{12}+\omega^{20})+\frac{1}{6}([\omega^{01},\omega^{12}]+[\omega^{12},\omega^{20}]+
[\omega^{20},\omega^{12}])+
\\
+\frac{1}{72}([[\omega^{01}+\omega^{12}+\omega^{20},\omega^{01}],\omega^{01}]+
[[\omega^{01}+\omega^{12}+\omega^{20},\omega^{12}],\omega^{12}]+
[[\omega^{01}+\omega^{12}+\omega^{20},\omega^{20}],\omega^{20}])+\OO(\omega^4)
\end{multline}

\subsubsection*{\bf Remark on divergencies}
Calculating values of 1-loop Feynman graphs for effective action on
simplex reduces essentially to calculating super-traces over
infinite-dimensional space of differential forms. These might
contain divergencies. As we have seen in section \ref{calculating S
for D=1}, this is not the case for dimension D=1: only finitely many
terms of the monodromy matrix (written in monomial basis) are
non-zero, and thus the super-trace is a sum of finitely many terms.

For dimension $D=2$ we also carried out a calculation of super-trace
for $q^{(2)}$ in monomial basis. For this case diagonal elements of
monodromy matrix do not vanish on monomials of high degree.
Moreover, super-traces of monodromy matrix on 0-forms and on 1-forms
diverge if calculated separately. If we employ the regularization
that is the monomial degree cut-off, i.e. we calculate super-trace
of monodromy acting on monomials of total degree $<N$, these
divergences are logarithmic: $\Str_{\Omega^0(\Delta^2)}\sim \log N$
and $\Str_{\Omega^1(\Delta^2)}\sim \log N$. But in the total
super-trace over all differential forms these divergencies cancel,
and the answer for $q^{(2)}$ is finite.

We also made a calculation of super-trace for $q^{(2)}$ in
``coordinate representation'', i.e. in basis of $\delta$-functions
of coordinates on simplex, centered in different points (thus
super-trace becomes an integral over the simplex). Here we also
encounter divergencies, and a nice way to handle them is to
introduce the following regularization: we change the Dupont's chain
homotopy operator $K$ to a regularized one $K_\e$, where $K_\e$ is
obtained by the same construction, described in section
\ref{Dupont's construction}, where we redefine the dilation map
$\phi_i$ to act on $[0,1-\e]\times\Delta^D$ instead of
$[0,1]\times\Delta^D$. Here $\e>0$ is an infinitesimal parameter.
This regularization immediately makes all answers in coordinate
repres
sult for $q^{(2)}$ coincides with
one obtained in monomial basis.

For case $D=3$ we calculated $q^{(2)}$ in coordinate representation
only (these calculations are technically simpler than in monomial
basis). In principle the corresponding super-trace could have not
just logarithmic, but even a linear (in cut-off parameter)
divergence. But, employing regularization $K\rightarrow K_\e$, we
obtain a finite answer.

\end{document}